\documentclass{aa}  
\usepackage{natbib}
\bibpunct{(}{)}{;}{a}{}{,}  
\usepackage{graphicx}
\usepackage{aalongtable}
\usepackage{txfonts}
\begin{document}
   \title{Variability and the X-ray/UV ratio of active galactic nuclei}

   \author{F. Vagnetti \inst{1}, S. Turriziani \inst{1}\fnmsep\thanks{Visitor at ASI Science Data Center, via Galileo Galilei, 00044 Frascati, Italy}, D. Trevese \inst{2}, M. Antonucci \inst{1} }

   \offprints{F. Vagnetti,\\ \email{fausto.vagnetti@roma2.infn.it}}

   \institute
       {Dipartimento di Fisica, Universit\`a di Roma ``Tor Vergata'', Via della Ricerca Scientifica 1, I-00133, Roma, Italy
       \and
       Dipartimento di Fisica, Universit\`a di Roma ``La Sapienza'', Piazzale Aldo Moro 2, I-00185 Roma, Italy}   

   \date{}

 
  \abstract
   {The observed relation between the X-ray radiation from active galactic nuclei, originating in the corona, and the optical/UV radiation from the disk is usually  described by the anticorrelation between the UV to X-ray slope $\alpha_{ox}$ and the UV luminosity. Many factors can affect this relation, including:  i) enhanced X-ray emission associated with the jets of radio-loud AGNs, ii) X-ray absorption associated with the UV broad absorption line (BAL) outflows, iii) other X-ray absorption not associated with BALs, iv) intrinsic X-ray weakness, v) UV and X-ray variability, and non-simultaneity of UV and X-ray observations. The separation of these effects provides information about the intrinsic $\alpha_{ox}-L_{UV}$ relation and its dispersion, constraining models of disk-corona coupling.}
   {We use simultaneous UV/X-ray observations to remove the influence of non-simultaneous measurements from the $\alpha_{ox}-L_{UV}$ relation.}
   {We extract simultaneous data from the second XMM-Newton serendipitous source catalogue (XMMSSC) and the XMM-Newton Optical Monitor Serendipitous UV Source Survey catalogue (XMMOMSUSS), and derive the single-epoch $\alpha_{ox}$ indices. We use ensemble structure functions to analyse multi-epoch data.}
   {We confirm the anticorrelation of $\alpha_{ox}$ with $L_{UV}$, and do not find any evidence of a dependence of $\alpha_{ox}$ on $z$. The dispersion in our simultaneous data ($\sigma\sim 0.12$) is not significantly smaller than in previous non-simultaneous studies, suggesting that ``artificial $\alpha_{ox}$ variability'' introduced by non-simultaneity is not the main cause of dispersion. ``Intrinsic $\alpha_{ox}$ variability'' , i.e., the true variability of the X-ray to optical ratio, is instead important, and accounts for $\sim 30\%$ of the total variance, or more. ``Inter-source dispersion", due to intrinsic differences in the average $\alpha_{ox}$ values from source to source, is also important. The dispersion introduced by variability is mostly caused by the long timescale variations, which are expected to be driven by the optical variations.}
{}

\authorrunning{F.Vagnetti et al.}
\titlerunning{X/UV ratio of AGNs}

   \maketitle

\section{Introduction}
The relationship between the X-ray and optical/UV luminosity of active galactic nuclei (AGNs) is usually described in terms of the index $\alpha_{ox}=0.3838 \log(L_X/L_{UV})$, i.e., the slope of a hypothetical power law between 2500 \AA\ and 2 keV rest-frame frequencies. The X-ray and UV monochromatic luminosities are correlated over 5 decades as $L_X \propto L_{UV}^k$, with $k\sim0.5-0.7$, and this provides an anticorrelation $\alpha_{ox} = a \log L_{UV}$ + const, with $-0.2\la a\la -0.1$ \citep[e.g.,][]{avni86,vign03,stra05,stef06,just07,gibs08}. One of the main results of these analyses is that QSOs are universally X-ray luminous and that X-ray weak QSOs are very rare \citep[e.g.,][]{avni86,gibs08}, but it is not yet known if the same is true for moderate luminosity AGNs. 
UV photons are generally believed to be radiated from the QSO accretion disk, while X-rays are supposed to originate in a hot coronal gas of unknown geometry and disk-covering fraction. The X-ray/UV ratio provides information about the balance between the accretion disk and the corona, which is not yet understood in detail. The $\alpha_{ox}-L_{UV}$ anticorrelation implies that AGNs redistribute their energy in the UV and X-ray bands depending on the overall luminosity, with more luminous AGNs emitting fewer X-rays per unit UV luminosity than less luminous AGNs \citep{stra05}. It has been proposed that the anticorrelation can be caused by
the larger dispersion in the luminosities in the UV than the X-ray band for a population with intrinsically uniform $\alpha_{ox}$ \citep{la-f95,yuan98}; however, more recent analyses based on samples with wider luminosity ranges confirm the reality of the relationship \citep{stra05}. \citet{gibs08} stressed the quite large scatter in the X-ray brightness of individual sources about the average relation and investigated the possible causes of the dispersion. Part of this scatter, usually removed \citep[e.g.,][]{stra05,stef06,just07,gibs08} is caused by radio-loud quasars, which are relatively X-ray bright because of the enhanced X-ray emission associated with their jets \citep[e.g.,][]{worr87}, and to broad absorption line (BAL) quasars, which are relatively X-ray faint \citep[e.g.,][]{bran00} due to X-ray absorption associated with the UV BAL outflows. 
Additional causes of deviations from the average $\alpha_{ox}-L_{UV}$ relation include: i) X-ray absorption not associated with BALs, ii) intrinsic X-ray weakness, and iii) UV and X-ray variability, possibly in association with non-simultaneous UV and X-ray observations. In particular, \citet{gibs08} estimate that variability may be responsible for 70\%-100\% of the $\alpha_{ox}$ dispersion, and that a few percent ($<2$\%) of all quasars are intrinsically X-ray weak by a factor of 10, compared to the average value at the same UV luminosity. A large fraction of intrinsically X-ray weak sources would suggest that coronae may frequently be absent or disrupted in QSOs. An extreme case is PHL 1811, which is X-ray weak by a factor $\sim$70, studied in detail by \citet{leig07}, who propose various scenarios, including disk/corona coupling by means of magnetic reconnections, Compton cooling of the corona by unusually soft optical/UV spectrum, and the photon trapping of X-ray photons and their advection to the black hole. The influence of variability on the $\alpha_{ox}-L_{UV}$ relation can be divided into two different effects: i) non-simultaneity of X-ray and UV measurements, which we call ``artificial $\alpha_{ox}$ variability'', and ii) true variability in the X-ray/UV ratio, which we refer to as ``intrinsic $\alpha_{ox}$ variability''. It is beneficial to analyse simultaneously acquired X-ray and UV data to eliminate the effect of the artificial variability and search for the intrinsic X-ray/UV ratio and/or its variability. On a rest-frame timescale of a few years, the optical/UV variability of QSOs has been estimated to be $\sim$30\% \citep[e.g.,][]{gial91,vand04}, while X-ray variability has been estimated to be $\sim$40\% for Seyfert 1 AGNs \citep{mark03}. On intermediate timescales, the relation between X-ray and optical/UV variability may be due to either: i) the reprocessing of X-rays into thermal optical emission, by means of irradiation and heating of the accretion disk, or ii) Compton up-scattering, in the hot corona, of optical photons emitted by the disk. In the former case, variations in the X-ray flux would lead optical/UV ones, and vice versa in the latter case. Cross-correlation analyses of X-ray and optical/UV light curves allow us to constrain models for the origin of variability. The main results obtained so far, on the basis of simultaneous X-ray and optical observations, indicate a cross-correlation between X-ray and UV/optical variation on the timescale of days, and in some cases delays of the UV ranging from 0.5 to 2 days have been measured \citep{smit07}. Simultaneous X-ray/UV data can be obtained by the XMM-Newton satellite, which carries the co-aligned Optical Monitor (OM). The second XMM-Newton serendipitous source catalogue (XMMSSC) \citep{wats09} is available online in the updated incremental version 2XMMi\footnote{http://heasarc.gsfc.nasa.gov/W3Browse/xmm-newton/xmmssc.html}. The XMM-Newton Optical Monitor Serendipitous UV Source Survey catalogue (XMMOMSUSS) is also available online\footnote{http://heasarc.gsfc.nasa.gov/W3Browse/xmm-newton/xmmomsuss.html}. We look for simultaneous measurements of the $\alpha_{ox}$ index from XMM/OM catalogues, to provide at least partial answers to the following questions: how large is the effect of non-simultaneous X-ray/UV observations on the dispersion about the average $\alpha_{ox}-L_{UV}$ relationship? Is there any spectral X-UV variability for individual objects? Do their $\alpha_{ox}$ harden in the bright phases or vice versa? Which constraints do these measurements place on the relationship between the accretion disk and the corona? 

The paper is organised as follows. Section 2 describes the data extracted from the archival catalogues. Section 3 describes the SEDs of the sources and the evaluation of the specific UV and X-ray luminosities. Section 4 discusses the $\alpha_{ox}-L_{UV}$ anticorrelation and its dispersion. In Sect. 5, we present the multi-epoch data and discuss the intrinsic X/UV variability of individual sources. Section 6 provides notes about individual peculiar sources. Section 7 discusses and summarises the results.

Throughout the paper, we adopt the cosmology H$_{o}$=70 km s$^{-1}$ Mpc$^{-1}$, $\Omega_{m}$=0.3, and $\Omega_{\Lambda}$=0.7.

\section{The data}
The updated incremental version 2XMMi of the second XMM-Newton serendipitous source catalogue (XMMSSC) \citep{wats09} is available online and contains 289083 detections between 2000 February 3 and 2008 March 28\footnote{After the submission of the article, a note has been distributed about ``Incorrect EPIC band-4 fluxes in the 2XMM and 2XMMi catalogues'' (XMM-Newton News \#105, http://xmm.esac.esa.int/external/xmm\_news/news\_list/). This affects 83 observations among the 315 in Table 1, which is corrected in agreement with the new data released by the XMM-Newton Survey Science Centre. All our analysis is also corrected with the new data.}. The net sky area covered by the catalogue fields is  $\sim 360$ deg$^2$. 

XMMOMSUSS is a catalogue of UV sources detected serendipitously by the Optical Monitor (OM) onboard the XMM-Newton observatory and is a partner resource to the 2XMM serendipitous X-ray source catalogue. The catalogue contains source detections drawn from 2417 XMM-OM observations in up to three broad-band UV filters, performed between 2000 February 24 and 2007 March 29. The net sky area covered is between 29 and 54 square degrees, depending on UV filter. The XMMOMSUSS catalogue contains 753578 UV source detections above a signal-to-noise ratio threshold limit of 3-$\sigma$, which relate to 624049 unique objects.

We first correlated the XMMSSC with the XMMOMSUSS catalogue to search for X-ray and UV sources with a maximum separation of 1.5 arcsec, corresponding to $\sim 1 \sigma$ uncertainty in the X-ray position. This yields 22061 matches. To obtain simultaneous X-ray and UV data, we searched for data from the same XMM-Newton observations, comparing the parameters OBS\_ID and OBSID of the XMMSSC and XMMOMSUSS catalogue, respectively, that identify uniquely the XMM-Newton pointings. This reduces the set to 8082 simultaneous observations. For the correlations, we used the Virtual Observatory application TOPCAT\footnote{http://www.star.bris.ac.uk/$\sim$mbt/topcat/}. 

We then correlated this table with the Sloan Digital Sky Survey (SDSS) Quasar Catalogue, Data Release 5, to provide optical classifications and redshifts for the matched objects \citep{schn07}. Using again a maximum distance of 1.5 arcsec  (uncertainty in the X-ray position), we found 310 matches. Increasing the maximum distance up to 5 arcsec, we add only 5 matches, none of which has a separation $>2$ arcsec. This indicates that, in spite of the relatively small (1.5 arcsec $\sim 1\sigma$)  cross-correlation radius adopted to reduce the contamination, the resulting incompleteness (at the present flux limit) is negligible.
The X-ray to optical ratios of the added 5 sources are not peculiar, therefore we used the entire sample of 315 matches. This also includes multi-epoch data for 46 sources (from 2 to 9 epochs each) and single-epoch observations for 195 more sources, for a total number of 241 sources. 

To estimate the probability of false identifications, we applied an arbitrary shift of 1 arcmin in declination to the X-ray coordinates of the 8082 simultaneous observations, and we found 219 UV/X-ray spurious associations, i.e., 2.7\%. This would correspond to $\sim 8$ spurious matches among the 315 observations of our final sample.

The relevant data of the sources are reported in Table 1, where:
Col. 1 corresponds to the source serial number;
Col. 2, the observation epoch serial number;
Col. 3  source name;
Col. 4  epoch (MJD);
Col. 5  redshift;
Col. 6  radio-loud flag (1=radio-loud, 0=radio-quiet, -1=unclassified);
Col. 7  BAL flag (1=BAL, 0=non-BAL);
Col. 8  log of the specific luminosity at 2500\AA\ in erg s$^{-1}$ Hz$^{-1}$;
Col. 9  log of the specific luminosity at 2 keV in erg s$^{-1}$ Hz$^{-1}$;
Col. 10 UV to X-ray power law index $\alpha_{ox}$;
Col. 11  residual of $\alpha_{ox}$ w.r.t. the adopted $\alpha_{ox}$-$L_{UV}$ correlation;
and Col. 12  hardness ratio between the bands 1-2 keV and 2-4.5 keV.

The sources span a region in the luminosity-redshift plane with $0.1\la z\la 3$ and $10^{29}$ erg s$^{-1}$ Hz$^{-1}  \la L_{UV} \la 10^{32}$ erg s$^{-1}$ Hz$^{-1}$, as shown in Fig. 1.

\begin{figure}
\centering
\resizebox{\hsize}{!}{\includegraphics{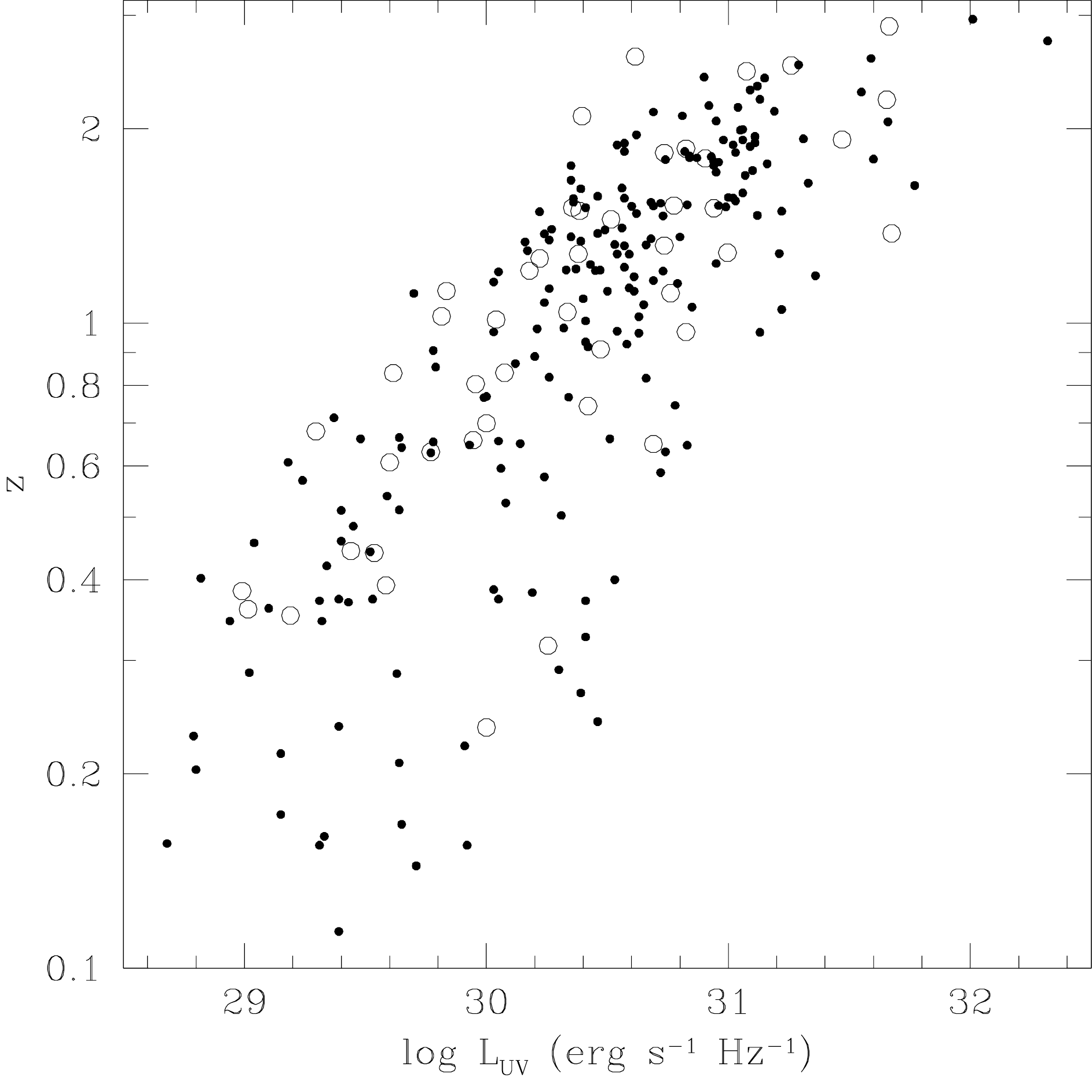}}
\caption{The sources in the luminosity-redshift plane. All the sources in Table 1 are reported. Small dots correspond to single-epoch data, while open circles indicate average luminosity values of multi-epoch sources.}
\end{figure}

\section{Evaluation of the specific luminosities}

\subsection{UV}
The Optical Monitor onboard XMM-Newton is described in detail in \citet{maso01}.  The set of filters included within the XMMOMSUSS catalogue is described in a dedicated page at MSSL\footnote{http://www.mssl.ucl.ac.uk/$\sim$mds/XMM-OM-SUSS/SourcePropertiesFilters.shtml.}. The filters are called UVW2, UVM2, UVW1, U, B, and V, with central wavelengths 1894\AA, 2205\AA, 2675\AA, 3275\AA, 4050\AA, and 5235\AA, respectively. The last three filters are similar, but not identical, to the Johnson UBV set. 

In the evaluation of the rest-frame luminosities, it is inadvisable to apply k-corrections using fixed power laws, because the local slope of the power law\footnote{We adopt spectral indices following the implicit sign convention, $L_\nu\propto\nu^{\alpha}$.} at the emission frequency corresponding to the observed bandpasses changes as a function of the source redshift, between $\sim -0.5$ and $\sim -2$ \citep[see, e.g.,][]{rich06}. The effective slope to compute specific luminosity at 2500\AA\ is an appropriate average of the slopes between the emission frequency and the frequency corresponding to 2500\AA.

One or more specific fluxes, up to six, are reported in XMMOMSUSS for the filters effectively used for each source, depending on observational limitations at each pointing. We were therefore able to compute optical-UV spectral energy distributions (SEDs) for each source. We derived specific luminosities at the different emission frequencies of the SEDs according to the classical formula

\begin{equation}
L_\nu(\nu_e)=F_\nu(\nu_o) {4\pi D_L^2\over 1+z}\, .
\end{equation}

\noindent
The result is plotted in Fig. 2, where SEDs with 2-6 frequency points are shown as lines, while small circles represent sources with only 1 frequency point. Black lines and circles refer to sources with data at a single epoch, while colours are used for multi-epoch sources. Data from the same source are plotted with the same colour, but more sources are represented with the same colour. The continuous curve covering the entire range of the plot is the average SED computed by \citet{rich06} for Type 1 quasars from the SDSS.

\begin{figure}
\centering
\resizebox{\hsize}{!}{\includegraphics{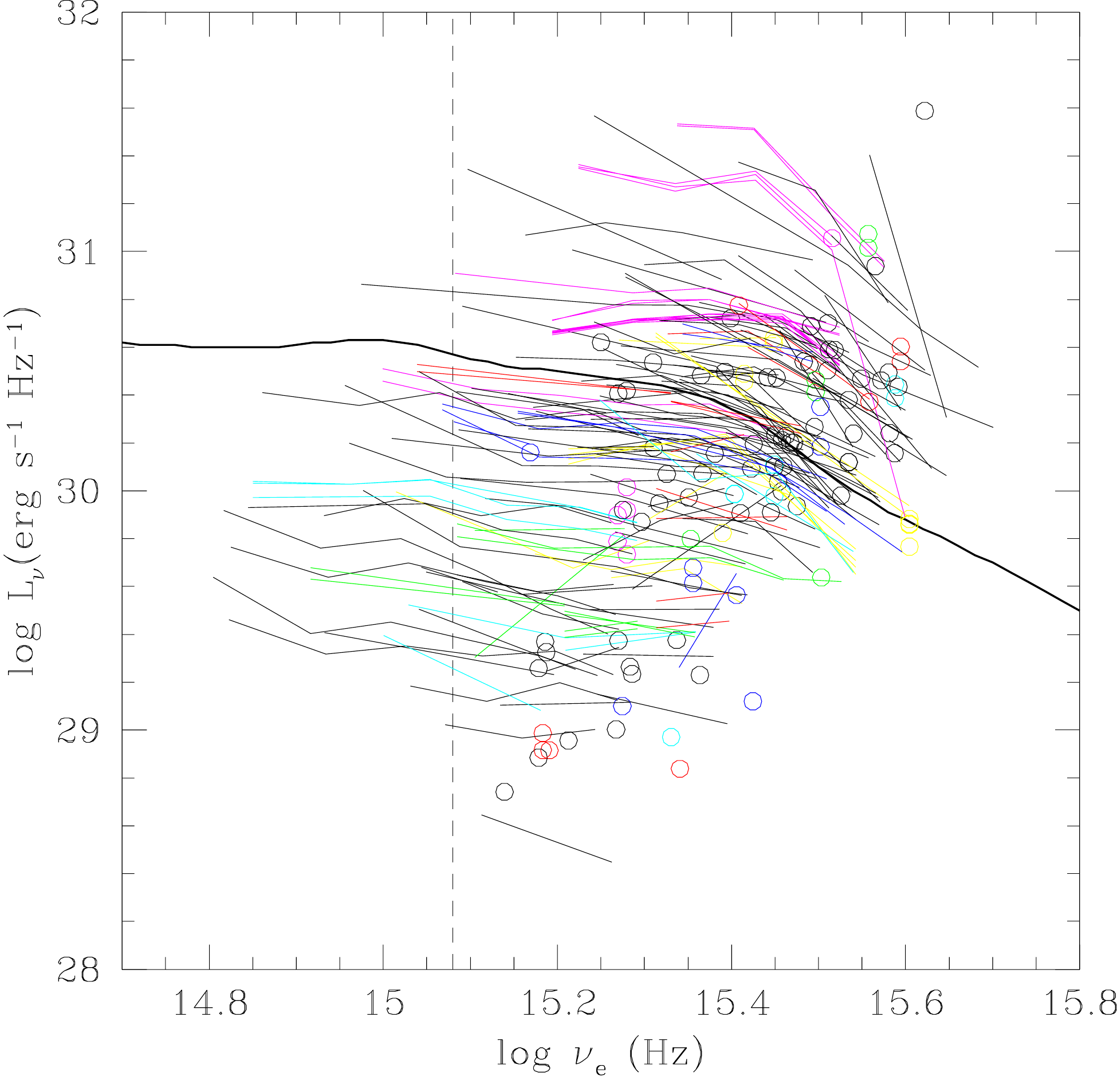}}
\caption{Spectral energy distributions from the available OM data. Sources with 2 or more frequency points are shown as lines, while small circles represent sources with only 1 frequency point. Black lines and circles refer to sources with data at a single epoch, while coloured data refer to multi-epoch sources. Data from the same source are plotted with the same colour, but more sources are represented with the same colour. The continuous curve covering all the range of the plot is the average SED computed by \citet{rich06} for Type 1 quasars from the SDSS.}
\end{figure}

The specific luminosity at 2500\AA , ($\log\nu_e=15.08$), called $L_{UV}$ for brevity, is evaluated as follows: i) if the SED of the source extends across a sufficiently wide range at low frequency, crossing the $\log\nu_e=15.08$ line (see Fig. 2), $L_{UV}$ is computed as an interpolation of the SED values in the 2 nearest frequency points; ii) in the other cases, i.e., if $\log\nu_e>15.08$ for all the SED, we use a curvilinear extrapolation, adopting the shape of the average SED by \citet{rich06}, shifting it vertically to match the specific luminosity of the source at the lowest frequency point available, say $\nu_1$, and applying a correction factor between $\log\nu_1$ and 15.08. Another possibility would be to extrapolate the source's SED using a power law with the same slope as that between the two lowest frequency points, but this is not applicable when there is only 1 frequency point, and is inappropriate when $\log\nu_1\ga 15.3$, because it is then in a region where the average SED by \citet{rich06} steepens. We therefore do not apply this power law extrapolation, and use instead the curvilinear extrapolation (ii) described above. However, we tested the use of this power  law extrapolation for the subset of SEDs for which it can be applied, and computed the $\alpha_{ox}-L_{UV}$ relation as described in the following (Sect. 4). We found similar slopes (within 0.010) and dispersions (within 0.005), which does not influence our final conclusions.

\subsection{X-ray}
X-ray fluxes are provided by the XMMSSC catalogue integrated in 5 basic energy bands, 0.2-0.5 keV (band 1), 0.5-1 keV (band 2), 1-2 keV (band 3), 2-4.5 keV (band 4), and 4.5-12 keV (band 5) \citep{wats09}. Power law distributions with photon index\footnote{With the usual convention of explicit minus sign for the photon index, $P(E)\propto E^{-\Gamma}$ and with the implicit sign adopted by us for the energy index $\alpha$, the relation between the two indices is $\Gamma=1-\alpha$.} $\Gamma=1.7$ and absorbing column density $N_H=3\times 10^{20}$ cm$^{-2}$ are assumed in the computation of the fluxes.

To evaluate the specific luminosity at 2 keV (which we call $L_X$ for brevity), we can use the flux in one of the two adjacent bands, 3 or 4. Since the fluxes are computed with negligible absorption, we prefer to use the band 4, which is less absorbed than the band 3 in type-2 obscured AGNs. It would also be possible to directly measure rest-frame 2 keV flux from observed low-energy bands 1 or 2, but - again - this would provide in some cases an absorbed flux. We therefore use the power law integral

\begin{equation}
F_X({\rm 2-4.5\,keV})=\int_{\rm 2\,keV}^{\rm 4.5\,keV} F_\nu({\rm 2\,keV})\left({\nu\over\nu_{\rm 2\,keV}}\right)^{1-\Gamma} d\nu\end{equation}
and determine the specific flux at 2 keV (observed frame) to be:

\begin{equation}
F_\nu({\rm 2\,keV})={F_X({\rm 2-4.5\,keV})\over \nu_{\rm 2\,keV}} {2-\Gamma\over 2.25^{2-\Gamma}-1}\quad .
\end{equation}

\noindent We then apply a standard power law k-correction

\begin{equation}
L_\nu({\rm 2\,keV})=F_\nu({\rm 2\,keV}) {4\pi D_L^2\over (1+z)^{2-\Gamma}}\quad ,
\end{equation}
adopting $\Gamma=1.7$ as assumed in the catalogue.

\section{The $\alpha_{ox}-L_{UV}$ anticorrelation}
We define, as usual

\begin{equation}
\alpha_{ox}={\log({L_{2{\rm\,keV}}/L_{2500{\rm\,\AA}}})\over \log({\nu_{2{\rm\,keV}}/\nu_{2500{\rm\,\AA}}})}=0.3838 \log\left({L_{2{\rm\,keV}}\over L_{2500{\rm\,\AA}}}\right)
\end{equation}

\noindent
and show in Fig. 3, $\alpha_{ox}$ as a function of $L_{UV}$ for all the sources in Table 1, including also multi-epoch measurements where available. Radio-loud quasars and BAL quasars are also shown with different symbols, and they are then removed from the main correlation.

\begin{figure}
\centering
\resizebox{\hsize}{!}{\includegraphics{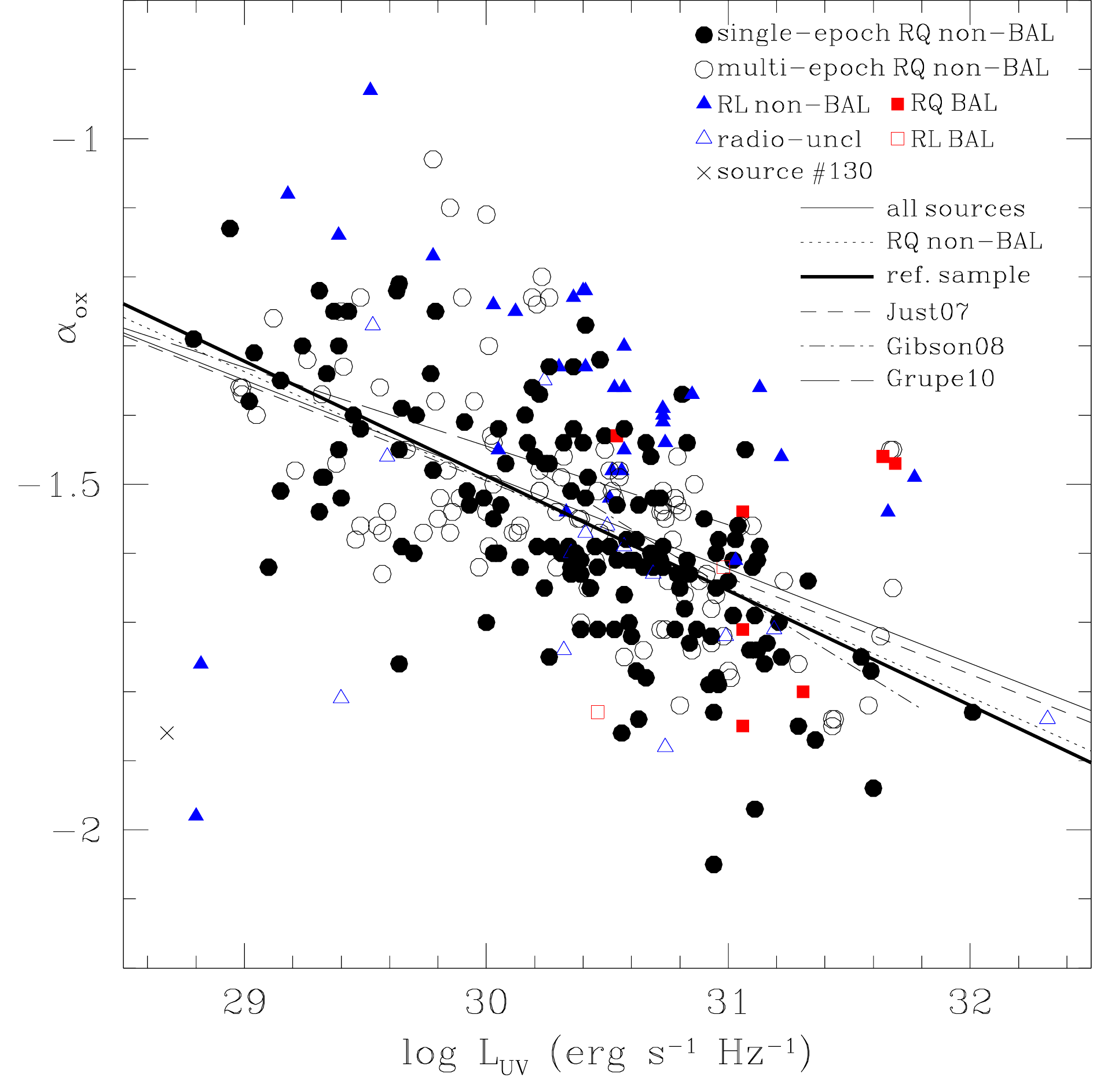}}
\caption{$\alpha_{ox}$ as a function of the 2500\AA\ specific luminosity $L_{UV}$, for all the 315 measurements of the sources in our sample, including multi-epoch measurements. Filled and open circles are, respectively, single-epoch and multi-epoch measurements of radio-quiet, non-BAL AGNs. BAL AGNs are plotted as filled (radio-quiet) and open (radio-loud) squares. Filled triangles represent radio-loud, non-BAL, sources, and open triangles are radio-unclassified sources. The $\times$ symbol indicates the anomalously X-ray-weak source \#130. Linear fits are represented as thin continuous (all the sources), dotted (radio-quiet non-BAL sources), thick continuous (excluding also source \#130). The short-dashed line is the best-fit reported by \citet{just07}. The dot-dashed, and long-dashed lines are the best-fit relations found by \citet{gibs08}, and by \citet{grup10}, respectively, and are plotted for limited ranges of luminosities, as analysed in the corresponding works.}
\end{figure}

Radio flux density at 1.4 GHz from FIRST radio survey \citep{beck95} is directly available in the SDSS-DR5 Quasar Catalog, where radio sources are associated with SDSS positions adopting a cross-correlation radius of 2 arcsec \citep{schn07}. In a few cases, additional radio information is taken from the NVSS survey \citep{cond98} and/or from the NASA Extragalactic Database (NED). In total, radio information is available for 228 sources out of 241 in Table 1. Following \citet{gibs08}, we assume a radio spectral index $\alpha=-0.8$ to estimate the specific luminosity at 5 GHz. We then calculate the radio-loudness parameter \citep[e.g.,][]{kell89},

\begin{equation}
R^*=L_\nu(5{\rm\,GHz})/L_{2500\AA}\, ,
\end{equation}

\noindent
and classify sources with $\log(R^*)\geq 1$ as radio-loud (RL), marking them with $f_{RL}=1$ in Table 1. Sources without detected radio flux or with $\log(R^*)< 1$ are classified as radio-quiet (RQ) and marked with $f_{RL}=0$. Sources without radio information from FIRST, NVSS, or NED are marked with $f_{RL}=-1$ .
Eight sources are present in the \citet{gibs08} and \citet{gibs09} catalogues as BAL quasars, and are accordingly marked in Table 1 with $f_{BAL}=1$.

As a first step, we show in Fig. 3 linear least squares fits corresponding to all the available measurements with the same weights, even for multi-epoch sources, as if they were different sources. The thin continuous line is a fit to all the sources, regardless of their radio-loudness and/or BAL characteristics, given by
\begin{equation}
\alpha_{ox} =(-0.137\pm0.013)\log L_{UV}+(2.610\pm0.401)\, .
\end{equation}

\noindent
A second fit, shown as a dotted line, corresponds to radio-quiet non-BAL sources, which are 193 of 241 in our sample
\begin{equation}
\alpha_{ox} =(-0.157\pm0.013)\log L_{UV}+(3.212\pm0.386)\, .
\end{equation}

\noindent
Radio-unclassified sources marked in Table 1 with $f_{RL}=-1$, are not included in this fit. Including them would make however a minor difference, as we have verified.

Most of the radio-loud sources in Fig. 3 are located above the fits, as expected, radio-loud quasars being known to have jet-linked X-ray emission components that generally lead to higher X-ray-to-optical ratios than those of radio-quiet quasars \citep[e.g.,][]{worr87}.

One source, \#130 in Table 1, appears to be very X-ray weak relative to the average correlation, as quantified in Sect. 4.1. This source is discussed further in Sect. 6 and we believe there are reasons to consider it to be anomalous. We then exclude it, so obtaining a reference sample of 192  radio-quiet non-BAL sources, not containing source \#130. We indicate with a thick continuous line the corresponding fit
\begin{equation}
\alpha_{ox} =(-0.166\pm0.012)\log L_{UV}+(3.489\pm0.377)\, .
\end{equation}

These correlations can be compared with that reported by \citet{gibs08}, which is shown in Fig. 3 as a dot-dashed line
\begin{equation}
\alpha_{ox} =(-0.217 \pm 0.036) \log L_{UV} + (5.075 \pm 1.118)\, .
\end{equation}

\noindent
and with those found by previous authors, usually flatter, as e.g. in \citet{just07}, whose fit is shown in Fig. 3 as a dashed line:
\begin{equation}
\alpha_{ox} =(-0.140 \pm 0.007) \log L_{UV} + (2.705 \pm 0.212)\, .
\end{equation}

The analysis of \citet{grup10} is also interesting, because it uses simultaneous X-ray and optical measurements from {\it Swift}, and has a yet flatter slope
\begin{equation}
\alpha_{ox} =(-0.114 \pm 0.014) \log L_{UV} + (1.177 \pm 0.305)\, .
\end{equation}

We note that the relations of \citet{gibs08} and \citet{grup10} are obtained by means of analyses in limited ranges of UV luminosities and redshifts, respectively of ($30.2<\log L_{UV}<31.8$, $1.7<z<2.7$) and ($26<\log L_{UV}<31$, $z<0.35$). This suggests a possible dependence of the slope of the $\alpha_{ox}-L_{UV}$ relation on luminosity or redshift or both, and is discussed further in Sect. 4.2.

We now limit ourselves to our reference sample of 192 sources, and show in Fig. 4 (as open circles) the average values of $L_{UV}$ and $\alpha_{ox}$ for 41 multi-epoch sources, together with the corresponding values for 151 single-epoch sources (black dots). Source \#45 is a known gravitational lens \citep{koch97}. \citet{char00} estimated that its luminosity is amplified by a factor $\sim 15$. We plot this source in Fig. 4 as an open square at the observed luminosity, and deamplified by a factor of 15 as an open circle, connected to the observed point by a dotted line. The parameter $\alpha_{ox}$ is not affected, as gravitational lensing is achromatic.

\begin{figure}
\centering
\resizebox{\hsize}{!}{\includegraphics{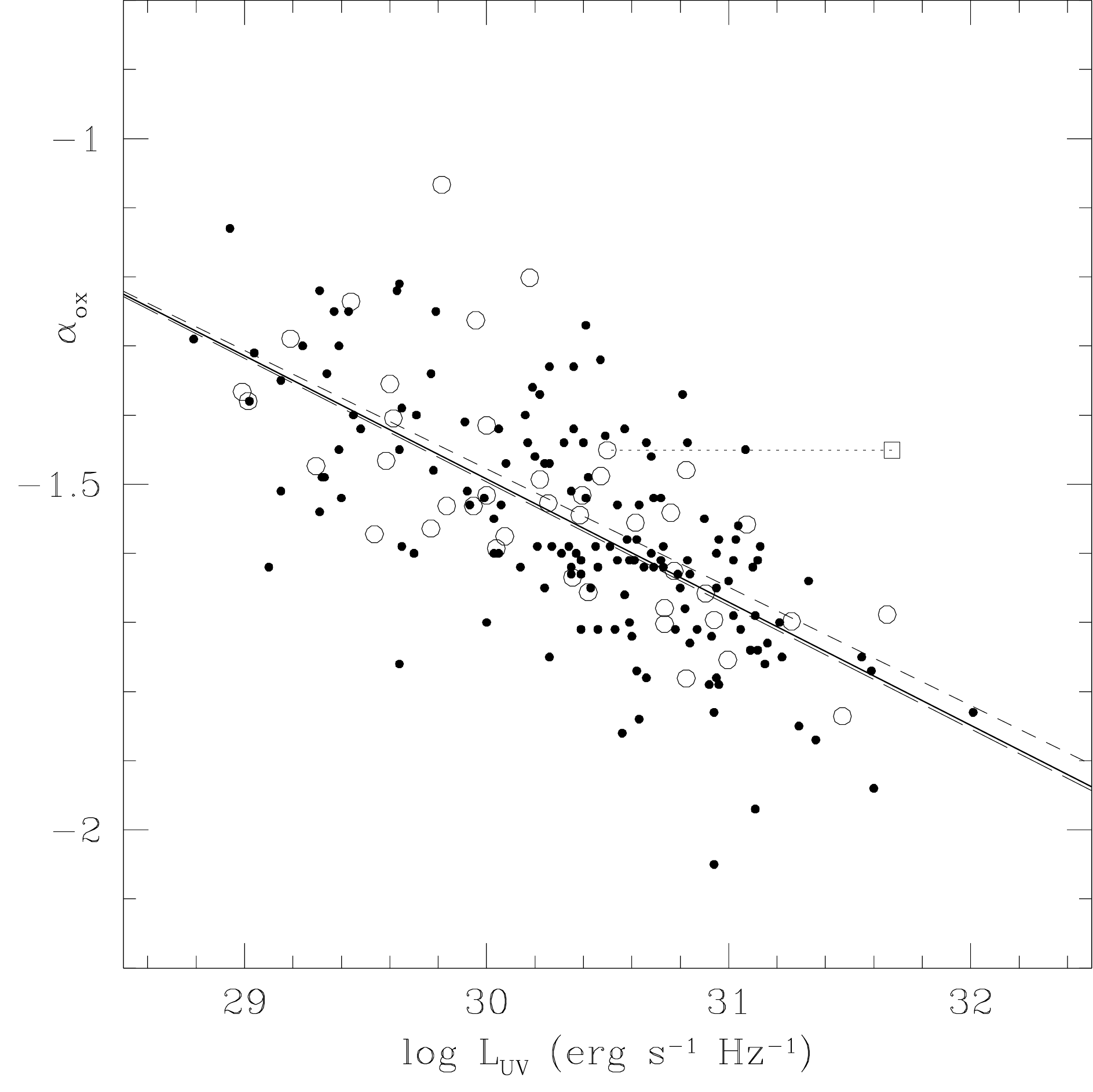}}
\caption{$\alpha_{ox}$ as a function of the 2500\AA\ specific luminosity $L_{UV}$, for the 192 radio-quiet non-BAL sources of the reference sample. Multi-epoch measurements of the same sources are averaged and shown as open circles, while black dots refer to single-epoch sources. Source \#45 is a gravitational lens and is shown with both its observed luminosity (as an open square) and its deamplified luminosity (as an open circle). The continuous line shows the least squares fit to the points. Dashed lines show separate fits to the single-epoch (long-dash) and multi-epoch (short-dash) sources.}
\end{figure}

The best-fit relation to the data in Fig. 4, including source \#45 with its deamplified luminosity, is
\begin{equation}
\alpha_{ox} =(-0.178\pm0.014)\log L_{UV}+(3.854\pm0.420)\, .
\end{equation}

\noindent
Separate fits for single-epoch and multi-epoch sources give, respectively, $\alpha_{ox} =(-0.179\pm0.016)\log L_{UV}+(3.863\pm0.482)$ and $\alpha_{ox} =(-0.171\pm0.029)\log L_{UV}+(3.657\pm0.877)$.

\subsection{Dispersion in $\alpha_{ox}$}

We adopt Eq. (13) as our reference $\alpha_{ox}(L_{UV})$ relation and investigate the dispersion of the sources around it. We therefore define the residuals
\begin{equation}
\Delta\alpha_{ox}=\alpha_{ox}-\alpha_{ox}(L_{UV})\, .
\end{equation}

We show in Fig. 5 the histograms of $\Delta\alpha_{ox}$, using the average values of multi-epoch measurements as in Fig. 4. Contour histogram represents all sources, the filled histogram the reference sample, and source \#130 is marked by a cross. The two histograms have standard deviations $\sigma=0.158$ and $\sigma=0.122$, respectively. The source \#130, with $\Delta\alpha_{ox}=-0.60$, differs by about $5\sigma$ from the reference relation, and appears X-ray weaker by a factor of $\sim 40$ than AGNs of the same UV luminosity.

\begin{figure}
\centering
\resizebox{\hsize}{!}{\includegraphics{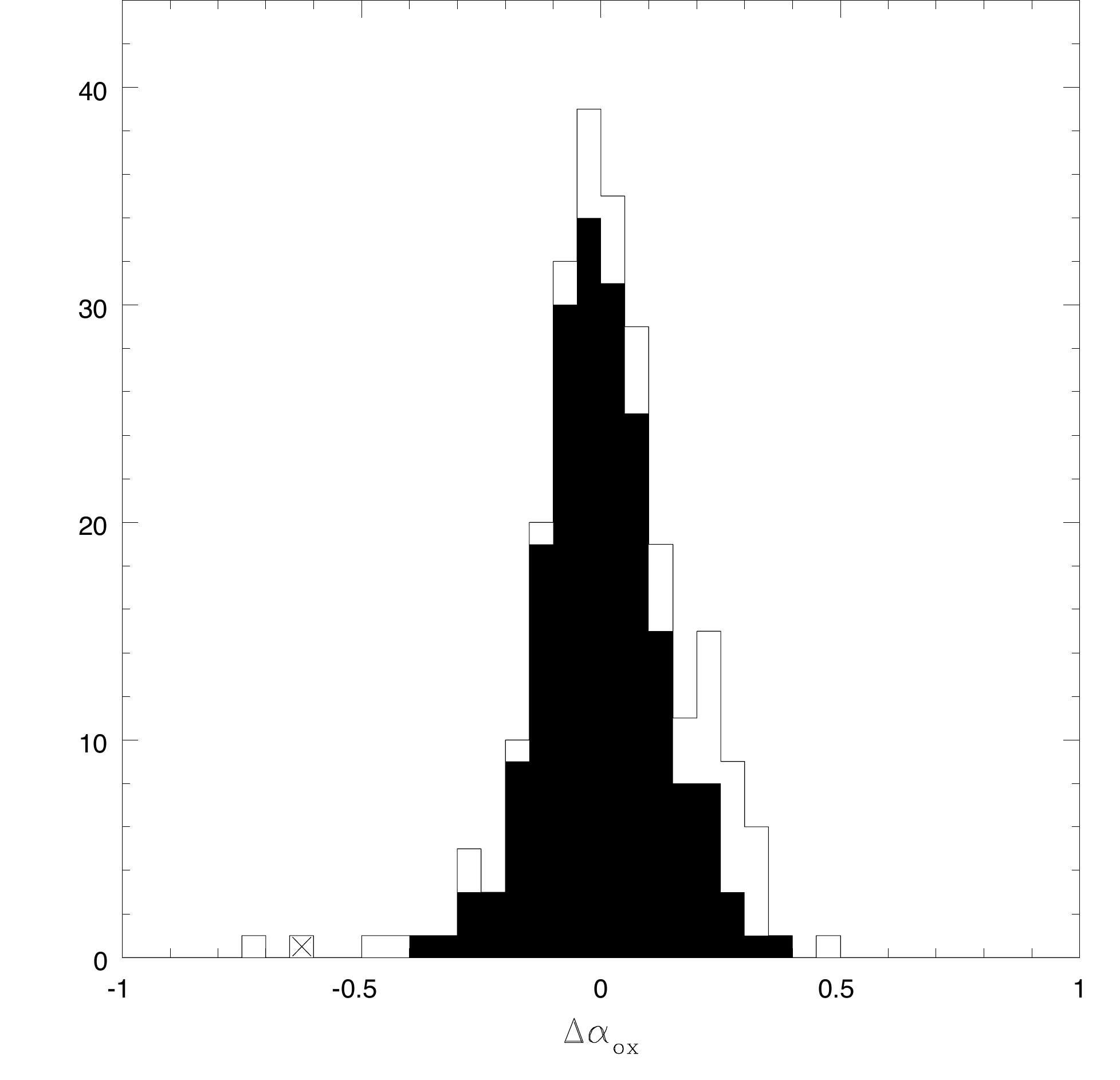}}
\caption{Histograms of the residuals $\Delta\alpha_{ox}$ (Eq. (14)) for all the sources (contour histogram), and for the reference sample (filled histogram). Source \#130 is marked by a cross. Dispersions for the two samples are $\sigma=0.158$ and $\sigma=0.122$, respectively.}
\end{figure}

The dispersion in our $\Delta\alpha_{ox}$ distribution is comparable to those obtained by, e.g., \citet{stra05}, \citet{just07}, and \cite{gibs08} on the basis of non-simultaneous X-ray and UV data, with values between 0.10 and 0.14. Our result based on simultaneous data eliminates a possible cause of dispersion due to ``artificial $\alpha_{ox}$ variability''. The dispersion is not lower than previous non-simultaneous estimates, thus it is probably caused by other factors affecting the X-ray/UV ratio. These could include: (i) ``intra-source dispersion'', caused by ``intrinsic $\alpha_{ox}$ variability'', i.e., true temporal change in the X-ray/UV ratio for individual sources,  and/or (ii) ``inter-source dispersion'', due to intrinsic differences in the average $\alpha_{ox}$ values from source to source, perhaps related to different conditions in the emitting regions.

\subsection{Dependence on $z$ and $L$}

To estimate the possible dependence of $\alpha_{ox}$ on redshift, we perform a partial correlation analysis, correlating $\alpha_{ox}$ with $L_{UV}$, with account for the effect of $z$, and correlating $\alpha_{ox}$ with $z$, taking account of the effect of $L_{UV}$.
For our reference sample of 192 radio-quiet, non-BAL, sources, we find a Pearson partial correlation coefficient $r_{\alpha L,z}=-0.51$, with a probability $P(>r)=1.3\times 10^{-12}$ for the null hypothesis that $\alpha_{ox}$ and $L_{UV}$ are uncorrelated. The other partial correlation coefficient is $r_{\alpha z,L}=0.05$ with $P(>r)=0.52$, which implies that there is no evidence of a correlation with $z$.

Our results agree with previous studies \citep{avni86,stra05,stef06,just07}, which also found no evidence of a dependence of $\alpha_{ox}$ on redshift \citep[see however][]{kell07}.

In the upper panel of Fig. 6, we plot the residuals $\alpha_{ox}-\alpha_{ox}(L_{UV})$, Eq. (14), as a function of $z$, which show no correlation ($r=0.027$,  $P(>r)=0.703$, $\Delta\alpha_{ox}=(0.005\pm0.014)z+(-0.006\pm0.018)$). In the lower panel, we plot the residuals $\alpha_{ox}-\alpha_{ox}(z)$ as a function of $\log L_{UV}$, after computing the average $\alpha_{ox}-z$ relation, $\alpha_{ox}(z)=(-0.139\pm0.016)z+(-1.394\pm0.022)$. These residuals clearly decrease with luminosity ($r=-0.305$,  $P(>r)=2.4\times 10^{-5}$, $\Delta\alpha_{ox}=(-0.067\pm0.015)\log L_{UV}+(2.050\pm0.465)$). Similar results were obtained by \citet{stef06}. These results suggest that the dependence of $\alpha_{ox}$ on $z$ is induced by the intrinsic dependence on $L_{UV}$ through the $L_{UV}-z$ correlation.

\begin{figure}
\centering
\resizebox{\hsize}{!}{\includegraphics{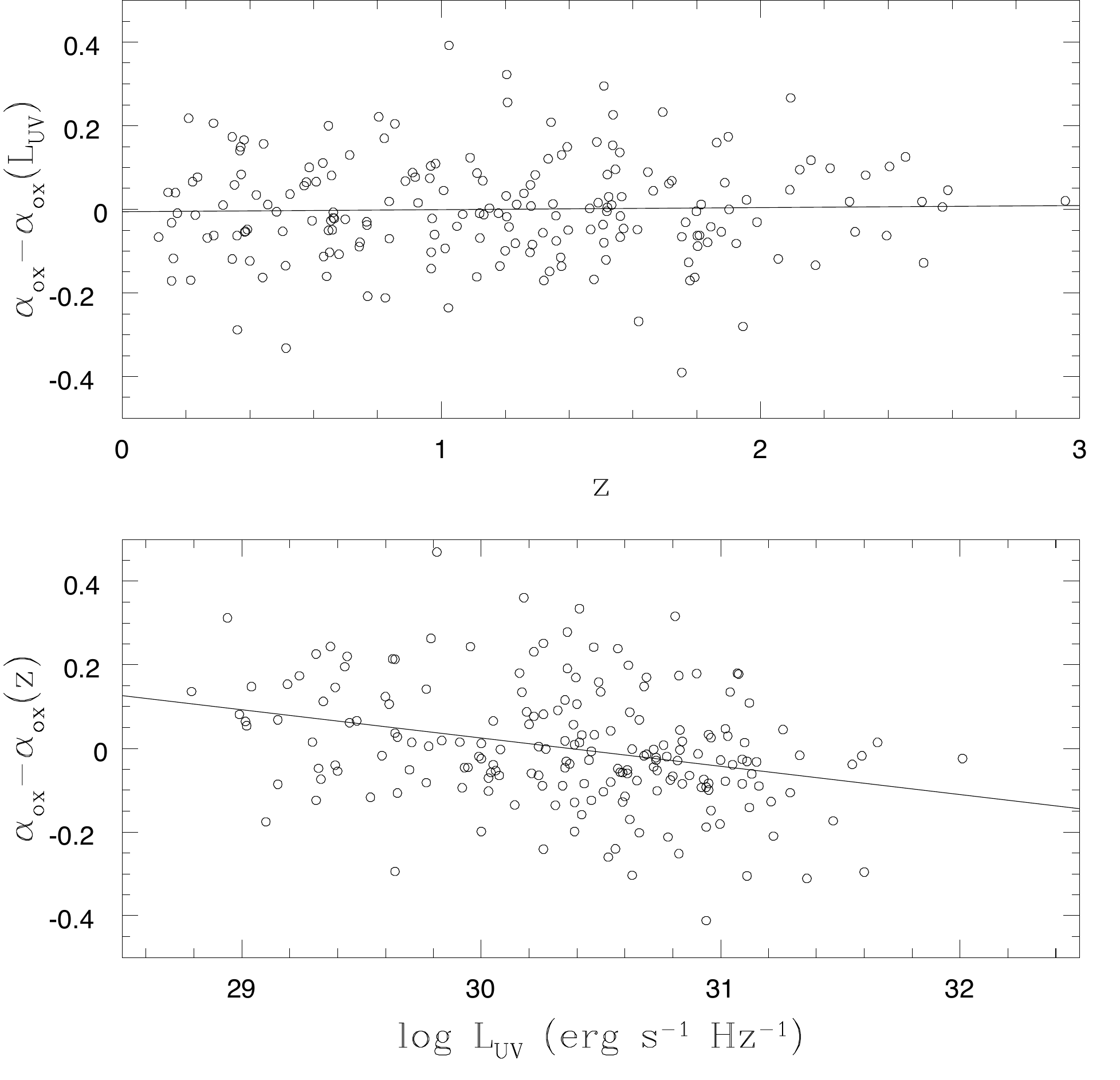}}
\caption{Upper panel: residuals $\alpha_{ox}-\alpha_{ox}(L_{UV})$, as a function of $z$. Lower panel: residuals $\alpha_{ox}-\alpha_{ox}(z)$ as a function of $\log L_{UV}$.}
\end{figure}

The slope of the $\alpha_{ox}-L_{UV}$ relation, according to the fits by \citet{gibs08} and \citet{grup10}, shown in Fig. 3 with the results of \citet{just07} and ourselves, may be flatter at lower luminosity and/or redshift. We divide our reference sample into two equally populated subsamples, $\log L_{UV}\lessgtr 30.43$, finding $\alpha_{ox} =(-0.137\pm0.029)\log L_{UV}+(2.639\pm0.878)$ for the low luminosity sources, and $\alpha_{ox} =(-0.193\pm0.038)\log L_{UV}+(4.319\pm1.182)$ for the high luminosity ones, while for the entire sample Eq. (13) is valid. A Student's-t test applied to the low-$L_{UV}$ and high-$L_{UV}$ subsamples gives a 12\% probability that they are drawn from the same parent distribution. A similar result was found by \citet{stef06}.

We similarly divide our sample into two redshift subsamples, $z\lessgtr 1.2$, finding $\alpha_{ox} =(-0.166\pm0.022)\log L_{UV}+(3.491\pm0.650)$ for the low $z$ sources, and $\alpha_{ox} =(-0.225\pm0.033)\log L_{UV}+(5.305\pm1.015)$ for the high $z$ sources. Application of the Student's-t test gives in this case a 7\% probability that low-$z$ and high-$z$ subsamples are drawn from the same parent distribution.

This suggests that the slope of the $\alpha_{ox}-L_{UV}$ relation may be $L_{UV}$- and/or $z$-dependent. However, the apparent dependence on $z$ can be an artifact of a true dependence on $L_{UV}$, or vice versa. A sample of sources evenly distributed in the $L-z$ plane is required to distinguish these dependences.

\section{Multi-epoch data}
We show in Fig. 7 the tracks of individual sources in the $\alpha_{ox}-L_{UV}$ plane, for the reference sample. Only 41 of 192 sources have multi-epoch information, and most of them exhibit small variations. Some sources (\#73, \#168) have strong variations in both $\alpha_{ox}$ and $L_{UV}$, but nearly parallel to the average $\alpha_{ox}-L_{UV}$ relation, therefore not contributing appreciably to the dispersion in $\Delta\alpha_{ox}$. A few sources (e.g., \#90, \#157, \#225) have appreciable or strong variations perpendicular to the average relation, possibly contributing to the overall dispersion. Figure 8 shows a histogram of the individual dispersions in $\Delta\alpha_{ox}$ for these 41 sources.

\begin{figure}
\centering
\resizebox{\hsize}{!}{\includegraphics{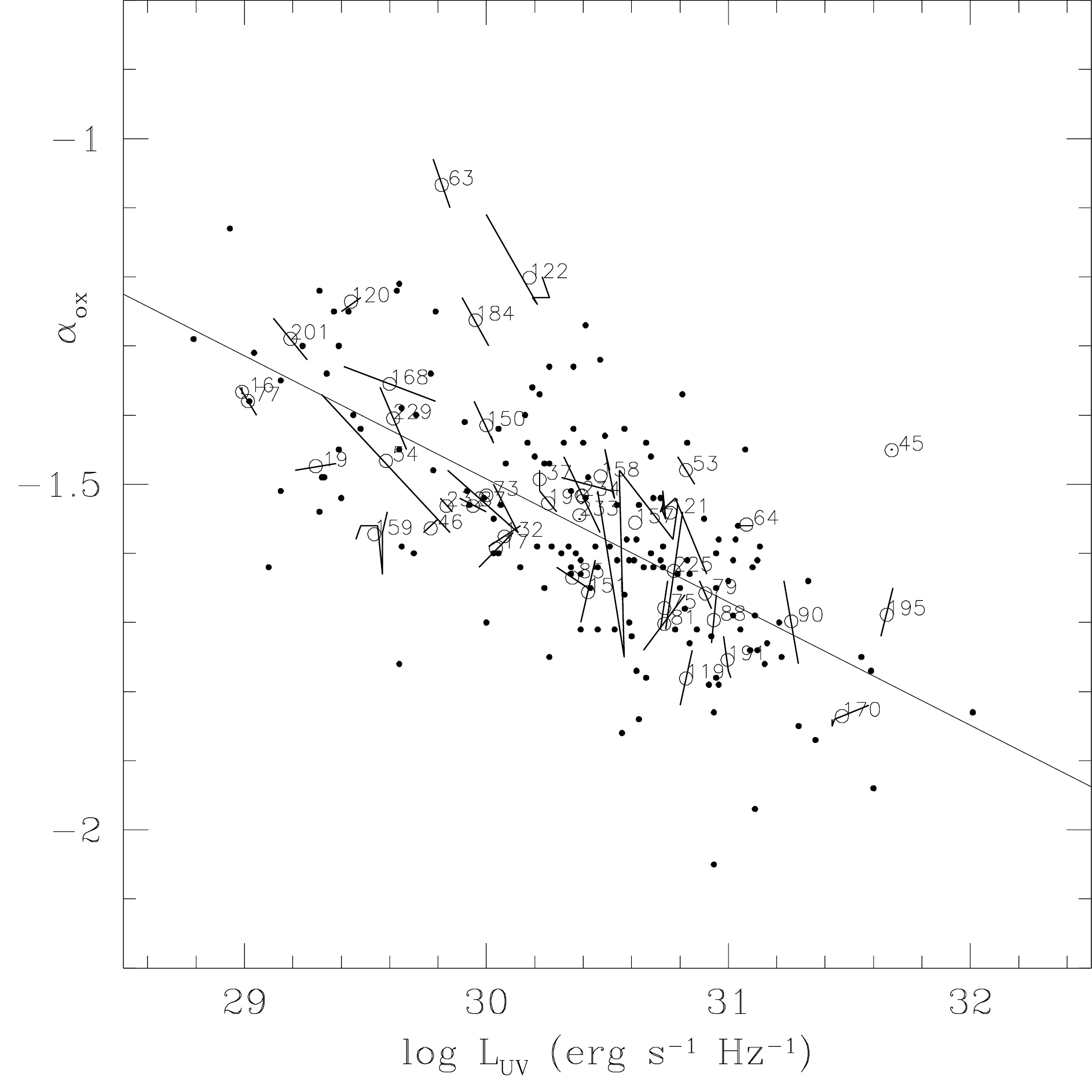}}
\caption{behaviour of individual radio-quiet non-BAL sources in the plane $\alpha_{ox}-L_{UV}$. Connected segments show the tracks of multi-epoch sources, while open circles represent the average values of the same sources, which are labeled with their serial numbers as in Table 1. Small dots refer to single-epoch sources. The straight line is the adopted $\alpha_{ox}-L_{UV}$ relation, Eq. (13).}
\end{figure}

\begin{figure}
\centering
\resizebox{\hsize}{!}{\includegraphics{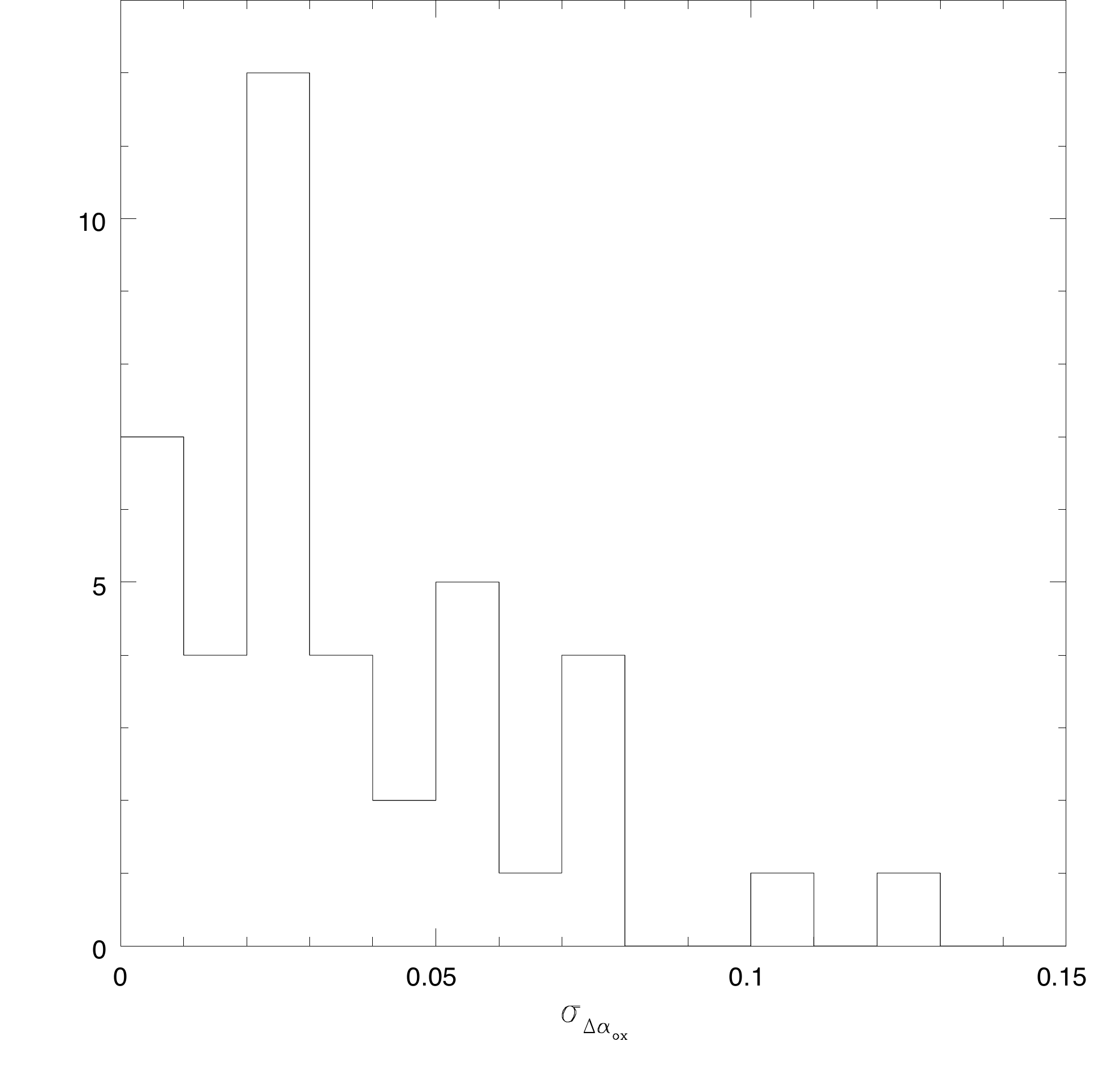}}
\caption{Histogram of the individual dispersions of $\Delta\alpha_{ox}$ for the 41 radio-quiet non-BAL sources with multi-epoch information.}
\end{figure}

Most sources have data at only 2 epochs, and only 9 sources have more epochs, up to 9. The individual variations occur on different timescales, from hrs to yrs, and cannot be directly compared with each other. It is however possible to build an ensemble structure function (SF) to describe the variability in a given quantity $A(t)$ for different rest-frame time-lags $\tau$. We define it as in \citet{dicl96}

\begin{equation}
SF(\tau)={\pi\over 2}\langle|A(t+\tau)-A(t)|\rangle\, .
\end{equation}

\noindent
The factor $\pi/2$ is introduced to measure SF in units of standard deviations, and  the angular brackets indicate the ensemble average over appropriate bins of time lag. The function $A(t)$ is usually a flux or luminosity in a given spectral band, or its logarithm. Here, we apply the definition of Eq. (15) to both $\alpha_{ox}(t)$ and the residuals $\Delta\alpha_{ox}(t)$. The result is illustrated in Fig. 9 for both functions. There is a clear increase in both SFs, which reach average variations of up to $\sim 0.07$ at $\sim 1$yr rest-frame. Unfortunately, the sampling is quite irregular, and most sources contribute with single points (corresponding to 2 epochs), while the few sources with more epochs have a greater weight in the ensemble statistic. To check whether the increase in SFs may be due to a single highly variable source, we compute new SFs by removing source \#157, which also has a relatively high number of epochs ($n=6$ epochs, therefore contributing $n(n-1)/2=15$ SF points), and find in this case a slightly  smaller (but still relevant) increase ($\sim 0.06$ at $\sim 1$yr).

These values can be compared with the dispersion in the residuals shown in Fig. 5, which is $\sigma=0.122$ for the reference sample. We note that the ensemble variability of $\Delta\alpha_{ox}$ was computed for only 41 multi-epoch sources, while the filled histogram shown in Fig. 5 also includes 151 single-epoch sources. We then checked whether the dispersions in the residuals for the single-epoch and multi-epoch subsamples are similar, being $\sigma=0.122$ and $\sigma=0.119$, respectively.

It then appears that variability in $\alpha_{ox}$ could account for a large part of the observed dispersion around the average $\alpha_{ox}-L_{UV}$ correlation. It is reasonable to expect that sources measured at single-epochs have temporal behaviours similar to those described by the SFs of Fig. 9, and that the variations in individual sources during their lifetime are similar to the variations measured from source to source at random epochs. However, the average temporal values of individual sources may differ, and ``inter-source dispersion'' may be present, in addition to ``intra-source dispersion'' (see Sect. 4.1). Assuming that other factors contributing to the dispersion can be neglected, the overall variance would then be:

\begin{equation}
\sigma^2=\sigma^2_{\rm intra-source}+\sigma^2_{\rm inter-source}\, .
\end{equation}

Our structure function analysis infers a value of 0.07 for the intra-source dispersion at 1 yr (or 0.06 if we remove the highly variable source \#157), while the total dispersion in the residuals shown in Fig. 5 is $\sigma\sim0.12$. This would indicate a $\sim 30\%$ contribution of intra-source dispersion to the total variance $\sigma^2$. However, the SF may increase further at longer time delays, so that the contribution of intra-source dispersion would be higher, while that of inter-source dispersion would be  constrained toward lower values.

Other factors may affect the dispersion, for example: (i) errors in the extrapolations of UV and X-ray luminosities, (ii) differences in galactic absorption, (iii) spurious inclusion of unknown BAL sources. From Fig. 2, it appears that a few sources have SEDs with anomalous slopes, and extrapolations with the average SED by \citet{rich06} infer in these cases poor luminosity estimates; however, this applies only to a small fraction of the sample. For X-rays, we adopted $\Gamma=1.7$ to be consistent with the fluxes catalogued in the XMMSSC; a distribution of $\Gamma$ values would introduce an extra dispersion. All these factors would probably contribute an additional term in Eq. (16). This would constrain more tightly the contribution of the inter-source dispersion, therefore increasing the relative weight of variability and intra-source dispersion.

A finer sampling of the SF and a homogeneous weight of the individual sources are however needed to quantify more definitely the contribution of variability, and which fraction has yet to be explained by other factors. Simultaneous UV and X-ray observations for a homogeneous sample of sources no greater than our own would be sufficient, assuming that each source is observed at $\sim 10$ epochs, spanning a monitoring time of a few years.

\begin{figure}
\centering
\resizebox{\hsize}{!}{\includegraphics{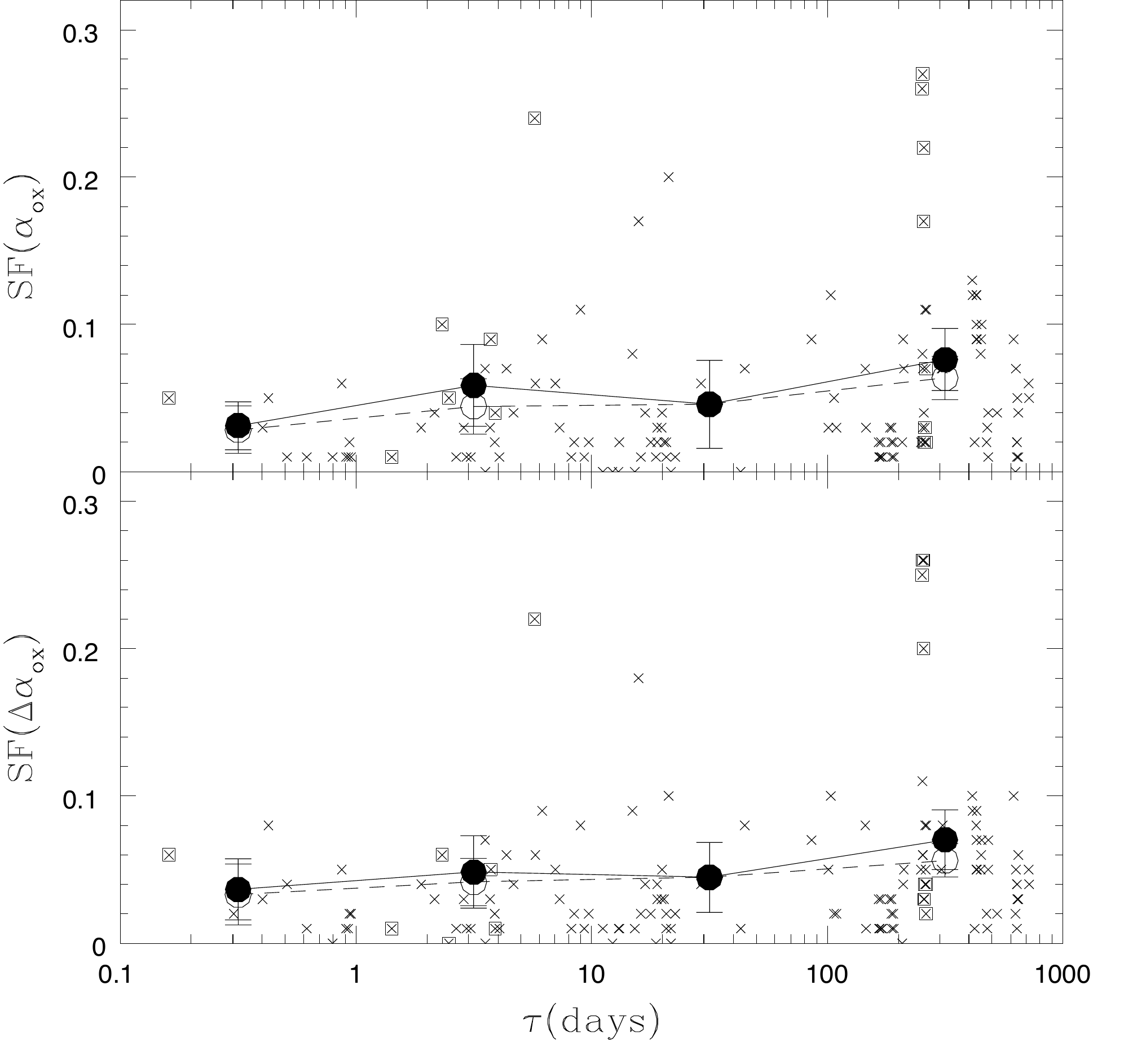}}
\caption{Rest-frame structure function of $\alpha_{ox}(t)$ (upper panel) and the residuals $\Delta\alpha_{ox}(t)$ (lower panel) for the 41 radio-quiet non-BAL sources with multi-epoch data. The crosses represent the contributions of the variations in individual sources. All the points marked by open squares refer to the variable source \#157. The filled circles connected by continuous lines represent the ensemble structure function for the set of 41 sources, in bins of $\Delta\log\tau=1$. The open circles connected by dashed lines correspond to the remaining set of 40 sources, after removing source \#157.}
\end{figure}

\section{Peculiar sources}
\subsection{2XMM J112611.6+425245}

We computed the X-ray luminosity and the $\alpha_{ox}$ spectral index starting from the X-ray flux in the 2-4.5 keV band (XMM-Newton band 4), as described in Sect. 3. Since 2XMM J112611.6+425245 (source \#130) is X-ray weak by a factor $\sim 40$, we analysed the X-ray information in the various XMM-Newton bands, available in the XMMSSC catalogue, and found this source to be even weaker in the softer 1-2 keV band (band 3), with a very high hardness ratio between the two bands, $HR3=(CR4-CR3)/(CR4+CR3)=0.52$, $CR3$ and $CR4$ being the count rates in the two bands. We then plot the sources of Table 1 in the plane $\Delta\alpha_{ox}-HR3$, to see whether X-ray weak sources are in some way related to particular values of the X-ray hardness ratio. This is shown in Fig. 10, where it can be seen that most sources are concentrated in a region with ``standard'' values around $\Delta\alpha_{ox}=0$ and $HR3\simeq -0.4$, while a few sources are located at greater distances, along tails in various directions. Source \#130, indicated by a $\times$ sign in the figure, is the most distant, and very X-ray weak and very hard.

\citet{hu08} report this source (which has a redshift $z=0.156$) in their study of the FeII emission in quasars, where it is shown that systematic inflow velocities of FeII emitting clouds are inversely correlated with Eddington ratios. The source 2XMM J112611.6+425245 has one of the highest measured inflow velocities, $v_{Fe}\sim 1700$ km s$^{-1}$. \citet{ferl09} argue about the high column densities, $N_H\sim 10^{22}-10^{23}$ cm$^{-2}$, necessary to account for the inflows in this class of quasars, and about the possibility that either UV or X-ray absorption are associated with the infalling component.

The source 2XMM J112611.6+425245 also has a high $HR4=(CR5-CR4)/(CR5+CR4)=0.63$, $CR5$ being the count rate in the 4.5-12 keV energy band. High values of $HR3$ and $HR4$ are used by \citet{nogu09} to select, on the basis of a modelling of the direct and scattered emission, a sample of AGNs hidden by geometrically thick tori. The hardness ratios of this source imply that it is a good candidate for that class of AGNs.

\begin{figure}
\centering
\resizebox{\hsize}{!}{\includegraphics{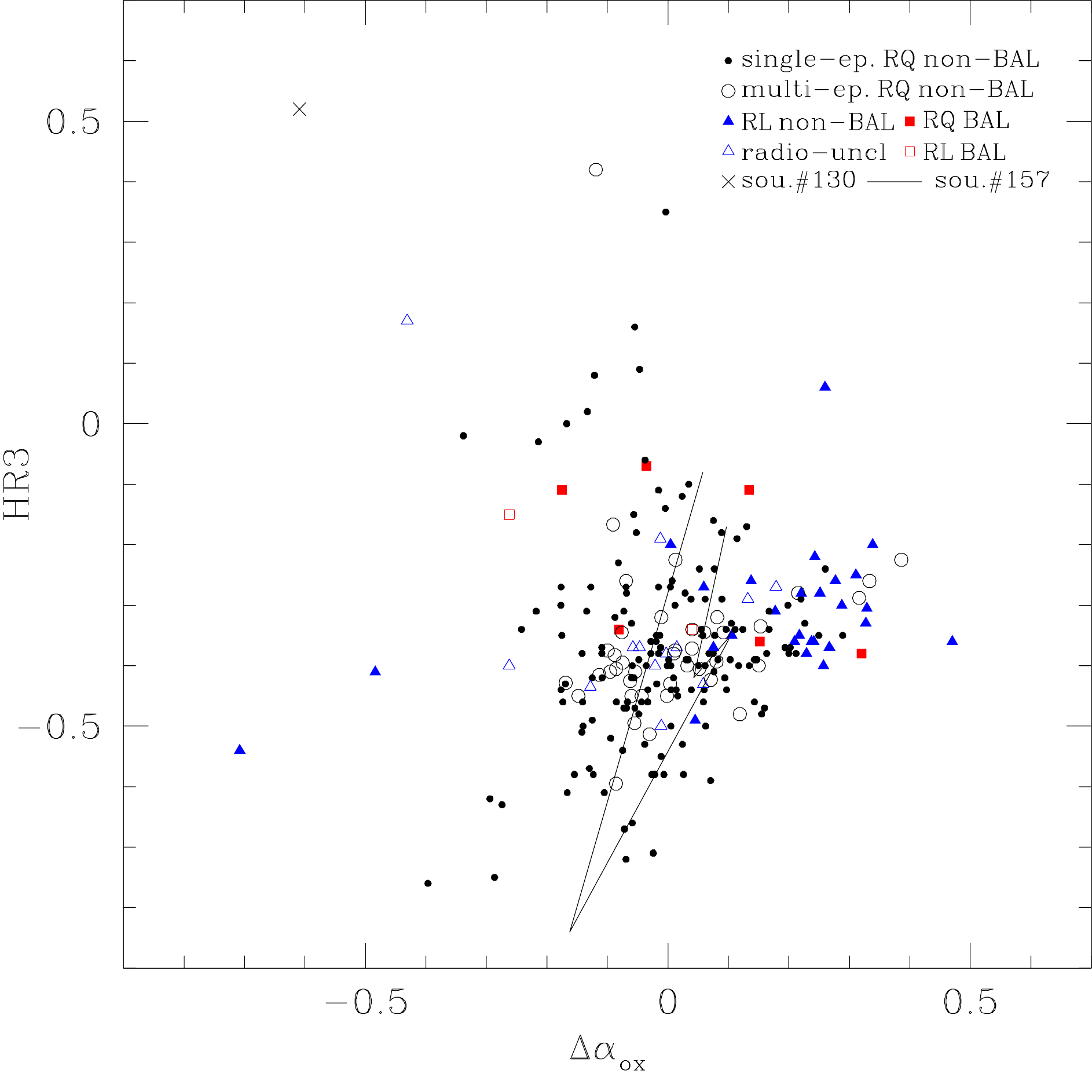}}
\caption{Plot of the sources of Table 1 in the plane $\Delta\alpha_{ox}-HR3$. Symbols as in Fig. 3. Sources with multi-epoch data are represented by their average values, except source \#157, whose strong variations are also shown by the connected segments.}
\end{figure}

\subsection{2XMM J123622.9+621526}
Source \#157 is one of those with the greatest variance in $\Delta\alpha_{ox}$. It exhibits even more extraordinary variations in $HR3$, which are shown in Fig. 10 by a broken line. It is in the Chandra Deep Field North, and its X-ray spectrum, analysed by \citet{baue04}, classifies it as an unobscured quasar. While its $UV$ luminosity has remained nearly constant, its $\Delta\alpha_{ox}$ and $HR3$ have varied by 0.22 and 0.76, respectively, between epochs 5 and 6 in Table 1, which differ by 20 days in the observed frame, i.e., less than a week in the rest-frame at the redshift $z=2.597$. 

\section{Discussion}

The behaviour of $\alpha_{ox}$, i.e., its dependence on luminosity and redshift, its dispersion and variability, are to be considered as symptoms of the relation between disk and corona emissions and their variabilities. 

It is generally believed that variable X-ray irradiation can drive optical variations by means of variable heating of the internal parts of the disk on relatively short timescales, days to weeks, while intrinsic disk instabilities in the outer parts of the disk dominate on longer timescales, months to years, propagating inwards and modulating X-ray variations in terms of Compton up-scattering in the corona \citep{czer04,arev06,arev09,papa08,mcha10}.

The structure functions of the light curves increase on long timescales both in the optical \citep[e.g.,][]{dicl96,vand04,baue09} and X-rays \citep[e.g.,][Vagnetti et al., in prep.]{fior98}. This, however, does not imply that the $\alpha_{ox}$ SF also increases with time lag. Larger changes (on long time lags) in both X-ray and UV fluxes may occur without changes in the spectral shape (i.e., with constant $\alpha_{ox}$). Our results shown in Fig. 9 indicate that this is not the case, i.e., that slope changes are indeed larger on longer timescales.

Moreover, it is evident from Fig. 9 that most of the dispersion about the $\alpha_{ox}-L_{UV}$ relation is due, in the present sample, to variations on timescales from months to years, which are associated with optically driven variations, according to the general belief.

The $\alpha_{ox}$ structure function does not distinguish between the hardening or softening of the optical to X-ray spectrum during brightening. This is instead described by the spectral variability parameter $\beta=\partial\alpha/\partial\log F$ \citep{trev01,trev02}, which can be adapted to the optical-X-ray case to become

\begin{equation}
\beta_{ox}={\delta\alpha_{ox}\over\delta\log L_{UV}}\,,
\end{equation}

\noindent
where $\beta_{ox}$ is the slope of the correlated variations $\delta\alpha_{ox}$ and $\delta L_{UV}$, and describes whether a source hardens when it brightens or vice versa, i.e., if the X-ray luminosity increases more than the optical or less. For example, single source variations parallel to the $\alpha_{ox}$ anticorrelation have a negative $\beta_{ox}$, while variations perpendicular to the correlation have $\beta_{ox}>0$. Both of these behaviours can be seen in Fig. 7. 

Of course, the different behaviour of the sources in the $\alpha_{ox}-L_{UV}$ plane may correspond to a different time sampling. 
Constraining physical models of the primary variability source and disk-corona coupling would require the analysis of $\beta_{ox}$ as a function of the time lag.
This analysis does not look feasible, i.e., to have statistical reliability, with the present sparse sampling. We can propose more conventional scenarios and note that since most of the variability, in the present sample, occurs on long ($\sim 1$ year ) timescale, it is presumably associated with optically driven variations. 
Considering all the measured variations $\delta\alpha_{ox}$ and $\delta L_{UV}$, we obtain the ``ensemble" average $\langle\beta_{ox}\rangle=-0.240$.
The negative sign implies that, on average, a spectral steepening occurs in the brighter phase. This is, in fact, consistent with larger variations in the UV band, driving the X-ray variability. 
The value of $\langle\beta_{ox}\rangle$ can be compared with the average slope of the $\alpha_{ox}-L_{UV}$ relation, Eq. (13), indicating that the UV excess in the brighter phase (steepening) is larger than the average UV excess in bright objects respect to faint ones.

Finally we emphasise that, despite its limitations, the present analysis illustrates the feasibility of an ensemble analysis of the $\alpha_{ox}-L_{UV}$ correlation, e.g., by considering the $\beta$ parameter as a function of time lag. What is presently missing is an adequate simultaneous X-ray-UV sampling, at relatively short time lags,  of a statistical AGN sample. An ensemble analysis may provide important constraints even when the total number of observations does not allow us to carry out  a cross-correlation analysis of X-ray and UV variations of individual sources.

We summarise our main results as follows:
\begin{itemize}
\item we have studied the $\alpha_{ox}-L_{UV}$ anticorrelation with simultaneous data extracted from the XMM-Newton Serendipitous Source catalogues;
\item we confirm the anticorrelation, with a slope (-0.178) slightly steeper w.r.t. \citet{just07};
\item we do not find evidence for a dependence of $\alpha_{ox}$ on redshift, in agreement with previous authors \citep[e.g.][]{avni86,stra05,stef06,just07};
\item there appears to be a flatter slope to the anticorrelation at low luminosities and low redshifts, in agreement with previous results by \citet{stef06};
\item the dispersion in our simultaneous data ($\sigma\sim 0.12$) is not significantly smaller w.r.t. previous non-simultaneous studies  \citep{stra05,just07,gibs08}, indicating that ``artificial $\alpha_{ox}$ variability'' introduced by non-simultaneity is not the main cause of dispersion;
\item ``intrinsic $\alpha_{ox}$ variability'' , i.e., true variability in the X-ray to optical ratio, is important, and accounts for $\sim 30\%$ of the total variance, or more;
\item ``inter-source dispersion", due to intrinsic differences in the average $\alpha_{ox}$ values from source to source, is also important;
\item the dispersion introduced by variability is mostly caused by the long timescale variations, which are expected to be dominated by the optical variations; the average spectral softening observed in the bright phase is consistent with this view;
\item distinguishing the trends produced by optical or X-ray variations may be achievable using the ensemble analysis of the spectral variability parameter $\beta_{ox}$ as a function of time lag; crucial information would be provided by wide field simultaneous UV and X-ray observations with relatively short (days-weeks) time lags.
\end{itemize}

\begin{acknowledgements}
We thank P. Giommi and A. Paggi for useful discussions.
This research has made use of the XMM-Newton Serendipitous Source Catalogue, which is a collaborative project involving the whole Science Survey Center Consortium.
This research has made use of the XMM-OM Serendipitous Ultra-violet Source Survey (XMMOMSUSS), which has been created at the University College London's (UCL's) Mullard Space Science Laboratory (MSSL) on behalf of ESA and is a partner resource to the 2XMM serendipitous X-ray source catalogue.
Funding for the SDSS and SDSS-II has been provided by the Alfred P. Sloan Foundation, the Participating Institutions, the National Science Foundation, the U.S. Department of Energy, the National Aeronautics and Space Administration, the Japanese Monbukagakusho, the Max Planck Society, and the Higher Education Funding Council for England. The SDSS Web Site is http://www.sdss.org/.
This research has made use of the NASA/IPAC Extragalactic Database (NED), which is operated by the Jet Propulsion Laboratory, California Institute of Technology, under contract with the National Aeronautics and Space Administration.
This work makes use of EURO-VO software, tools or services. The EURO-VO has been funded by the European Commission through contract numbers RI031675 (DCA) and 011892 (VO-TECH) under the 6th Framework Programme and contract number 212104 (AIDA) under the 7th Framework Programme.
S.T. acknowledges financial support through Grant ASI I/088/06/0.
\end{acknowledgements}

\bibliographystyle{aa}
\bibliography{14320.bib}{}

\onltab{1}{
\begin{longtable}{rccccrrccrrr}
\caption{The sources.}\\
\hline\hline
$N_{sou}$ &$N_{epo}$  &source  &epoch &$z$  &$f_{RL}^{\rm a}$  &$f_{BAL}^{\rm b}$  &$\log L_{UV}$ &$\log L_X$  &$\alpha_{ox}$ &$\Delta\alpha_{ox}$  &$HR3$  \\
(1) &(2)  &(3)  &(4) &(5)  &(6)  &(7)  &(8) &(9) &(10) &(11) &(12)  \\
\hline
\endfirsthead
\caption{continued.}\\
\hline\hline
$N_{sou}$ &$N_{epo}$  &source  &epoch &$z$  &$f_{RL}^{\rm a}$  &$f_{BAL}^{\rm b}$  &$\log L_{UV}$ &$\log L_X$  &$\alpha_{ox}$ &$\Delta\alpha_{ox}$  &$HR3$  \\
(1) &(2)  &(3)  &(4) &(5)  &(6)  &(7)  &(8) &(9) &(10) &(11) &(12)  \\
\hline
\endhead
\hline
\endfoot
  1 &   & 2XMM J003922.3+005951 & 53365.828  & 1.989 &  0 & 0   & 31.05  & 26.58  & $-$1.71  & $-$0.04  & $-$0.46 \\
  2 &   & 2XMM J010647.9+004628 & 52835.359  & 1.877 &  0 & 0   & 31.09  & 26.55  & $-$1.74  & $-$0.06  & $-$0.42 \\
  3 &   & 2XMM J011902.9$-$005633 & 52831.922  & 1.614 &  0 & 0   & 30.39  & 26.20  & $-$1.61  & $-$0.05  & $-$0.47 \\
  4 &   & 2XMM J014251.7+133352 & 51916.445  & 1.075 & $-$1 & 0   & 30.24  & 26.71  & $-$1.35  &  0.18  & $-$0.27 \\
  5 &   & 2XMM J014814.0+140853 & 52662.000  & 0.373 & $-$1 & 0   & 29.53  & 26.22  & $-$1.27  &  0.13  & $-$0.29 \\
  6 &   & 2XMM J015254.0+010435 & 52276.062  & 0.570 &  0 & 0   & 29.24  & 25.84  & $-$1.30  &  0.05  & $-$0.40 \\
  7 &   & 2XMM J015258.6+010507 & 52276.062  & 0.647 &  0 & 0   & 29.93  & 25.94  & $-$1.53  & $-$0.06  & $-$0.15 \\
  8 &   & 2XMM J015704.1$-$005656 & 53565.277  & 1.779 &  0 & 0   & 30.94  & 26.16  & $-$1.83  & $-$0.18  & $-$0.44 \\
  9 &   & 2XMM J015733.8$-$004823 & 53565.277  & 1.545 &  0 & 0   & 31.03  & 26.92  & $-$1.58  &  0.09  & $-$0.29 \\
 10 &   & 2XMM J020118.6$-$091936 & 53023.926  & 0.661 &  0 & 0   & 30.51  & 26.35  & $-$1.59  & $-$0.01  & $-$0.35 \\
 11 &   & 2XMM J021100.8$-$095138 & 53016.961  & 0.767 &  0 & 0   & 30.34  & 26.21  & $-$1.59  & $-$0.04  & $-$0.46 \\
 12 &   & 2XMM J023057.3$-$010033 & 53398.395  & 0.650 &  0 & 0   & 30.14  & 25.93  & $-$1.62  & $-$0.11  & $-$0.38 \\
 13 &   & 2XMM J024040.8$-$081309 & 53747.141  & 1.844 &  0 & 0   & 30.82  & 26.43  & $-$1.68  & $-$0.05  & $-$0.39 \\
 14 &   & 2XMM J024055.8$-$081952 & 53747.141  & 1.802 &  0 & 0   & 30.87  & 26.40  & $-$1.71  & $-$0.07  & $-$0.27 \\
 15 &   & 2XMM J024105.8$-$081153 & 53747.141  & 0.979 &  0 & 0   & 30.21  & 26.06  & $-$1.59  & $-$0.07  & $-$0.46 \\
 16 & 1 & 2XMM J024207.2+000038 & 51754.270  & 0.384 &  0 & 0   & 28.99  & 25.43  & $-$1.37  & $-$0.06  & $-$0.44 \\
 16 & 2 & 2XMM J024207.2+000038 & 51755.121  & 0.384 &  0 & 0   & 28.99  & 25.43  & $-$1.36  & $-$0.05  & $-$0.46 \\
 17 & 1 & 2XMM J024215.0$-$000209 & 51754.270  & 1.012 &  0 & 0   & 29.97  & 25.75  & $-$1.62  & $-$0.14  & $-$0.58 \\
 17 & 2 & 2XMM J024215.0$-$000209 & 51755.121  & 1.012 &  0 & 0   & 30.11  & 26.03  & $-$1.57  & $-$0.06  & $-$0.17 \\
 18 &   & 2XMM J024304.6+000005 & 51754.270  & 1.995 &  0 & 1   & 31.06  & 27.04  & $-$1.54  &  0.13  & $-$0.11 \\
 19 & 1 & 2XMM J024308.1$-$000126 & 51754.270  & 0.679 &  0 & 0   & 29.21  & 25.35  & $-$1.48  & $-$0.13  & $-$0.52 \\
 19 & 2 & 2XMM J024308.1$-$000126 & 51755.121  & 0.679 &  0 & 0   & 29.38  & 25.56  & $-$1.47  & $-$0.09  & $-$0.31 \\
 20 &   & 2XMM J025301.5$-$011148 & 52836.367  & 0.769 &  0 & 0   & 30.00  & 25.57  & $-$1.70  & $-$0.21  & $-$0.03 \\
 21 &   & 2XMM J030357.4$-$010906 & 53785.020  & 1.520 &  0 & 0   & 30.96  & 26.85  & $-$1.58  &  0.08  & $-$0.24 \\
 22 &   & 2XMM J030627.5$-$001816 & 53212.285  & 1.538 &  0 & 0   & 30.36  & 26.88  & $-$1.33  &  0.22  & $-$0.29 \\
 23 &   & 2XMM J030639.6+000725 & 52681.762  & 2.172 &  0 & 0   & 30.92  & 26.26  & $-$1.79  & $-$0.14  & $-$0.50 \\
 24 &   & 2XMM J030641.7+000109 & 52681.762  & 1.397 &  0 & 0   & 30.27  & 26.13  & $-$1.59  & $-$0.06  & $-$0.42 \\
 25 &   & 2XMM J030707.3$-$000424 & 52681.762  & 0.664 &  0 & 0   & 29.64  & 25.87  & $-$1.45  & $-$0.03  & $-$0.38 \\
 26 &   & 2XMM J033810.1+002324 & 52327.367  & 1.120 & $-$1 & 0   & 30.50  & 26.44  & $-$1.56  &  0.02  & $-$0.37 \\
 27 &   & 2XMM J033852.8+001905 & 52327.367  & 0.459 & $-$1 & 0   & 29.40  & 24.69  & $-$1.81  & $-$0.43  &  0.17 \\
 28 &   & 2XMM J034131.1$-$011405 & 53026.754  & 1.791 & $-$1 & 0   & 30.74  & 25.85  & $-$1.88  & $-$0.26  & $-$0.40 \\
 29 &   & 2XMM J073601.4+434455 & 52205.023  & 1.814 &  0 & 0   & 30.84  & 26.59  & $-$1.63  &  0.01  & $-$0.44 \\
 30 &   & 2XMM J074110.6+311200 & 52018.582  & 0.631 &  1 & 0   & 30.74  & 26.99  & $-$1.44  &  0.18  & $-$0.31 \\
 31 &   & 2XMM J074222.3+494147 & 52025.754  & 0.927 &  0 & 0   & 30.58  & 26.45  & $-$1.58  &  0.01  & $-$0.42 \\
 32 &   & 2XMM J080711.0+390419 & 52764.121  & 0.369 &  0 & 0   & 29.43  & 26.17  & $-$1.25  &  0.13  & $-$0.40 \\
 33 &   & 2XMM J081014.5+280337 & 52552.309  & 0.821 &  0 & 0   & 30.66  & 26.91  & $-$1.44  &  0.16  & $-$0.38 \\
 34 &   & 2XMM J081030.2+281326 & 52552.309  & 0.887 &  0 & 0   & 30.20  & 26.39  & $-$1.46  &  0.06  & $-$0.40 \\
 35 &   & 2XMM J081108.6+280500 & 52552.309  & 1.560 &  0 & 0   & 30.57  & 26.25  & $-$1.66  & $-$0.07  & $-$0.31 \\
 36 &   & 2XMM J082257.6+404149 & 53464.184  & 0.865 &  1 & 0   & 30.12  & 26.87  & $-$1.25  &  0.26  & $-$0.40 \\
 37 & 1 & 2XMM J084905.0+445714 & 52197.344  & 1.259 &  0 & 0   & 30.22  & 26.37  & $-$1.48  &  0.05  & $-$0.35 \\
 37 & 2 & 2XMM J084905.0+445714 & 52203.820  & 1.259 &  0 & 0   & 30.22  & 26.29  & $-$1.51  &  0.02  & $-$0.45 \\
 38 &   & 2XMM J085522.9+375425 & 53653.344  & 2.296 &  0 & 0   & 31.09  & 26.56  & $-$1.74  & $-$0.06  & $-$0.33 \\
 39 &   & 2XMM J085551.1+375752 & 53653.344  & 1.929 &  0 & 1   & 31.31  & 26.62  & $-$1.80  & $-$0.08  & $-$0.34 \\
 40 &   & 2XMM J085609.4+374928 & 53653.344  & 2.570 &  0 & 0   & 31.59  & 26.96  & $-$1.77  & $-$0.00  & $-$0.40 \\
 41 &   & 2XMM J085724.0+090349 & 52743.902  & 1.049 &  0 & 0   & 31.22  & 26.65  & $-$1.75  & $-$0.05  &  0.09 \\
 42 &   & 2XMM J085808.9+274522 & 53469.824  & 1.090 &  0 & 0   & 30.40  & 26.65  & $-$1.44  &  0.12  & $-$0.40 \\
 43 &   & 2XMM J090029.0+390145 & 53108.977  & 0.964 &  0 & 0   & 30.63  & 26.65  & $-$1.53  &  0.07  & $-$0.38 \\
 44 &   & 2XMM J091029.0+542719 & 53457.977  & 0.526 &  0 & 0   & 30.08  & 26.25  & $-$1.47  &  0.03  & $-$0.39 \\
 45 & 1 & 2XMM J091301.0+525929 & 52746.621  & 1.377 &  0 & 0   & 31.67  & 27.90  & $-$1.45  &  0.33  & $-$0.25 \\
 45 & 2 & 2XMM J091301.0+525929 & 52777.559  & 1.377 &  0 & 0   & 31.68  & 27.89  & $-$1.45  &  0.34  & $-$0.27 \\
 46 & 1 & 2XMM J091302.8+530322 & 52746.621  & 0.631 &  0 & 0   & 29.80  & 25.75  & $-$1.55  & $-$0.10  &  0.41 \\
 46 & 2 & 2XMM J091302.8+530322 & 52777.559  & 0.631 &  0 & 0   & 29.74  & 25.64  & $-$1.57  & $-$0.13  &  0.43 \\
 47 &   & 2XMM J091345.5+405629 & 52756.375  & 0.442 &  1 & 0   & 29.52  & 27.11  & $-$0.93  &  0.47  & $-$0.36 \\
 48 &   & 2XMM J091528.7+441633 & 53289.238  & 1.489 &  1 & 0   & 31.22  & 27.42  & $-$1.46  &  0.24  & $-$0.22 \\
 49 &   & 2XMM J091617.4+303038 & 52751.641  & 0.215 &  0 & 0   & 29.15  & 25.22  & $-$1.51  & $-$0.18  & $-$0.35 \\
 50 &   & 2XMM J092138.4+301546 & 52752.086  & 1.590 &  0 & 1   & 31.06  & 26.24  & $-$1.85  & $-$0.18  & $-$0.11 \\
 51 &   & 2XMM J092238.3+512120 & 53651.457  & 1.753 &  0 & 0   & 30.35  & 26.12  & $-$1.62  & $-$0.07  & $-$0.67 \\
 52 &   & 2XMM J092246.9+512037 & 53651.457  & 0.160 &  0 & 0   & 29.33  & 25.44  & $-$1.49  & $-$0.12  & $-$0.58 \\
 53 & 1 & 2XMM J093359.2+551550 & 52374.898  & 1.863 &  0 & 0   & 30.79  & 26.99  & $-$1.46  &  0.17  & $-$0.31 \\
 53 & 2 & 2XMM J093359.2+551550 & 52381.078  & 1.863 &  0 & 0   & 30.86  & 26.95  & $-$1.50  &  0.14  & $-$0.36 \\
 54 & 1 & 2XMM J094404.3+480647 & 53292.855  & 0.392 &  0 & 0   & 29.85  & 25.78  & $-$1.57  & $-$0.11  & $-$0.35 \\
 54 & 2 & 2XMM J094404.3+480647 & 53322.395  & 0.392 &  0 & 0   & 29.32  & 25.75  & $-$1.37  & $-$0.01  & $-$0.47 \\
 55 &   & 2XMM J094409.6+480813 & 53322.395  & 1.111 &  0 & 0   & 29.70  & 25.53  & $-$1.60  & $-$0.17  & $-$0.00 \\
 56 &   & 2XMM J094437.9+035936 & 53143.621  & 1.335 &  0 & 0   & 30.16  & 26.50  & $-$1.40  &  0.11  & $-$0.19 \\
 57 &   & 2XMM J094439.8+034940 & 53143.621  & 0.155 &  0 & 0   & 29.31  & 25.30  & $-$1.54  & $-$0.18  & $-$0.30 \\
 58 &   & 2XMM J095253.7+075040 & 52402.926  & 1.468 &  0 & 0   & 31.12  & 26.59  & $-$1.74  & $-$0.05  &  0.16 \\
 59 &   & 2XMM J095815.5+014922 & 53500.598  & 1.509 &  0 & 0   & 30.41  & 27.10  & $-$1.27  &  0.29  & $-$0.35 \\
 60 &   & 2XMM J095819.8+022903 & 53312.266  & 0.345 &  0 & 0   & 29.32  & 25.43  & $-$1.49  & $-$0.13  & $-$0.42 \\
 61 &   & 2XMM J095821.6+024628 & 53348.320  & 1.403 &  1 & 0   & 30.56  & 26.70  & $-$1.48  &  0.11  & $-$0.35 \\
 62 &   & 2XMM J095834.0+024427 & 53348.320  & 1.888 &  0 & 0   & 31.02  & 26.83  & $-$1.61  &  0.06  & $-$0.35 \\
 63 & 1 & 2XMM J095857.3+021314 & 52984.953  & 1.024 &  0 & 0   & 29.78  & 27.09  & $-$1.03  &  0.42  & $-$0.29 \\
 63 & 2 & 2XMM J095857.3+021314 & 53499.809  & 1.024 &  0 & 0   & 29.85  & 26.98  & $-$1.10  &  0.36  & $-$0.16 \\
 64 & 1 & 2XMM J095858.6+020139 & 53351.918  & 2.454 &  0 & 0   & 31.05  & 26.99  & $-$1.56  &  0.11  & $-$0.49 \\
 64 & 2 & 2XMM J095858.6+020139 & 53500.195  & 2.454 &  0 & 0   & 31.10  & 27.04  & $-$1.56  &  0.12  & $-$0.47 \\
 65 &   & 2XMM J095902.7+021906 & 52984.953  & 0.345 &  0 & 0   & 28.94  & 26.00  & $-$1.13  &  0.17  & $-$0.34 \\
 66 &   & 2XMM J095908.3+024309 & 53701.762  & 1.317 &  1 & 0   & 30.57  & 27.19  & $-$1.30  &  0.29  & $-$0.30 \\
 67 &   & 2XMM J095918.7+020951 & 52984.953  & 1.157 &  1 & 0   & 30.03  & 26.81  & $-$1.24  &  0.25  & $-$0.28 \\
 68 &   & 2XMM J095924.4+015954 & 53351.918  & 1.236 &  0 & 0   & 30.95  & 26.65  & $-$1.65  &  0.01  & $-$0.40 \\
 69 &   & 2XMM J095946.0+024743 & 53701.762  & 1.067 &  0 & 0   & 30.65  & 26.42  & $-$1.62  & $-$0.02  & $-$0.43 \\
 70 &   & 2XMM J095949.4+020141 & 53351.918  & 1.753 &  0 & 0   & 30.94  & 25.61  & $-$2.05  & $-$0.40  & $-$0.76 \\
 71 &   & 2XMM J095958.0+014327 & 53328.664  & 1.618 &  0 & 0   & 30.56  & 25.72  & $-$1.86  & $-$0.27  & $-$0.63 \\
 72 &   & 2XMM J100001.3+024845 & 53340.984  & 0.766 &  0 & 0   & 29.99  & 26.04  & $-$1.52  & $-$0.04  & $-$0.40 \\
 73 & 1 & 2XMM J100012.9+023522 & 52981.777  & 0.699 &  0 & 0   & 30.03  & 26.12  & $-$1.50  & $-$0.01  & $-$0.42 \\
 73 & 2 & 2XMM J100012.9+023522 & 53340.984  & 0.699 &  0 & 0   & 30.13  & 26.04  & $-$1.57  & $-$0.06  & $-$0.56 \\
 73 & 3 & 2XMM J100012.9+023522 & 53697.477  & 0.699 &  0 & 0   & 29.84  & 25.99  & $-$1.48  & $-$0.02  & $-$0.56 \\
 74 &   & 2XMM J100024.3+015053 & 52983.094  & 1.664 &  0 & 0   & 30.35  & 26.42  & $-$1.51  &  0.04  & $-$0.29 \\
 75 & 1 & 2XMM J100024.6+023148 & 52981.777  & 1.318 &  0 & 0   & 30.75  & 26.47  & $-$1.64  & $-$0.02  & $-$0.40 \\
 75 & 2 & 2XMM J100024.6+023148 & 53697.477  & 1.318 &  0 & 0   & 30.72  & 26.25  & $-$1.71  & $-$0.10  & $-$0.45 \\
 76 &   & 2XMM J100025.2+015852 & 52983.094  & 0.373 &  0 & 0   & 29.39  & 26.00  & $-$1.30  &  0.08  & $-$0.35 \\
 77 & 1 & 2XMM J100043.1+020637 & 52983.484  & 0.360 &  0 & 0   & 28.98  & 25.43  & $-$1.36  & $-$0.06  & $-$0.41 \\
 77 & 2 & 2XMM J100043.1+020637 & 53697.738  & 0.360 &  0 & 0   & 29.05  & 25.41  & $-$1.40  & $-$0.08  & $-$0.11 \\
 78 &   & 2XMM J100058.8+015359 & 53329.047  & 1.559 &  0 & 0   & 30.36  & 26.65  & $-$1.42  &  0.13  & $-$0.17 \\
 79 & 1 & 2XMM J100114.3+022356 & 52979.078  & 1.799 &  0 & 0   & 30.93  & 26.56  & $-$1.68  & $-$0.03  & $-$0.40 \\
 79 & 2 & 2XMM J100114.3+022356 & 53697.227  & 1.799 &  0 & 0   & 30.88  & 26.61  & $-$1.64  &  0.00  & $-$0.24 \\
 80 &   & 2XMM J100116.7+014053 & 53331.000  & 2.055 &  0 & 0   & 30.95  & 26.30  & $-$1.78  & $-$0.12  & $-$0.49 \\
 81 & 1 & 2XMM J100120.2+023341 & 52979.078  & 1.834 &  0 & 0   & 30.65  & 26.11  & $-$1.74  & $-$0.14  & $-$0.39 \\
 81 & 2 & 2XMM J100120.2+023341 & 53697.227  & 1.834 &  0 & 0   & 30.82  & 26.49  & $-$1.66  & $-$0.03  & $-$0.42 \\
 82 &   & 2XMM J100130.3+014304 & 53331.000  & 1.571 &  0 & 0   & 30.46  & 26.25  & $-$1.62  & $-$0.05  & $-$0.18 \\
 83 &   & 2XMM J100132.2+013419 & 53331.000  & 1.360 &  0 & 0   & 30.35  & 26.09  & $-$1.63  & $-$0.08  & $-$0.23 \\
 84 &   & 2XMM J100205.2+554258 & 52926.145  & 1.151 &  0 & 0   & 30.79  & 26.55  & $-$1.63  & $-$0.00  &  0.35 \\
 85 & 1 & 2XMM J100219.5+015537 & 53330.234  & 1.509 &  0 & 0   & 30.29  & 26.06  & $-$1.62  & $-$0.08  & $-$0.57 \\
 85 & 2 & 2XMM J100219.5+015537 & 53696.145  & 1.509 &  0 & 0   & 30.42  & 26.13  & $-$1.65  & $-$0.09  & $-$0.62 \\
 86 &   & 2XMM J100226.3+021923 & 53504.160  & 1.294 &  0 & 0   & 30.17  & 26.41  & $-$1.44  &  0.08  & $-$0.41 \\
 87 & 1 & 2XMM J100232.1+023537 & 53350.574  & 0.658 &  0 & 0   & 30.00  & 25.98  & $-$1.54  & $-$0.05  & $-$0.47 \\
 87 & 2 & 2XMM J100232.1+023537 & 53695.887  & 0.658 &  0 & 0   & 29.89  & 25.93  & $-$1.52  & $-$0.05  & $-$0.52 \\
 88 & 1 & 2XMM J100234.3+015011 & 53330.234  & 1.506 &  0 & 0   & 30.93  & 26.43  & $-$1.73  & $-$0.08  & $-$0.52 \\
 88 & 2 & 2XMM J100234.3+015011 & 53693.703  & 1.506 &  0 & 0   & 30.95  & 26.61  & $-$1.66  & $-$0.00  & $-$0.38 \\
 89 &   & 2XMM J100236.6+015949 & 53330.234  & 1.516 &  0 & 0   & 30.60  & 26.11  & $-$1.72  & $-$0.13  & $-$0.27 \\
 90 & 1 & 2XMM J100238.2+013747 & 53330.621  & 2.506 &  0 & 0   & 31.23  & 26.96  & $-$1.64  &  0.06  & $-$0.21 \\
 90 & 2 & 2XMM J100238.2+013747 & 53693.703  & 2.506 &  0 & 0   & 31.29  & 26.71  & $-$1.76  & $-$0.04  & $-$0.54 \\
 91 &   & 2XMM J100248.9+325130 & 53677.328  & 1.537 &  0 & 0   & 30.68  & 26.87  & $-$1.46  &  0.15  & $-$0.39 \\
 92 &   & 2XMM J100325.0+325305 & 53677.328  & 2.511 &  0 & 0   & 31.29  & 26.47  & $-$1.85  & $-$0.13  & $-$0.31 \\
 93 &   & 2XMM J100441.0+410944 & 53115.141  & 1.022 &  0 & 0   & 30.63  & 25.84  & $-$1.84  & $-$0.24  & $-$0.34 \\
 94 &   & 2XMM J100717.2+124543 & 52763.820  & 1.281 &  0 & 0   & 31.21  & 26.79  & $-$1.70  &  0.00  & $-$0.39 \\
 95 &   & 2XMM J100726.0+124856 & 52763.820  & 0.241 &  1 & 1   & 30.46  & 25.70  & $-$1.83  & $-$0.26  & $-$0.15 \\
 96 &   & 2XMM J101148.9+554102 & 52224.977  & 1.533 &  0 & 0   & 30.72  & 26.51  & $-$1.61  &  0.00  & $-$0.27 \\
 97 &   & 2XMM J101720.6+385738 & 52215.426  & 0.629 &  0 & 0   & 29.77  & 26.27  & $-$1.34  &  0.11  & $-$0.33 \\
 98 &   & 2XMM J101850.4+411508 & 52216.008  & 0.577 &  0 & 0   & 30.24  & 26.41  & $-$1.47  &  0.06  & $-$0.46 \\
 99 &   & 2XMM J101857.5+412549 & 52216.008  & 2.123 &  0 & 0   & 30.69  & 26.73  & $-$1.52  &  0.09  & $-$0.18 \\
100 &   & 2XMM J102003.7+081837 & 52055.531  & 2.094 &  0 & 0   & 30.81  & 27.24  & $-$1.37  &  0.26  & $-$0.24 \\
101 &   & 2XMM J102117.7+131546 & 52764.246  & 1.565 &  0 & 0   & 31.00  & 26.73  & $-$1.64  &  0.02  & $-$0.53 \\
102 &   & 2XMM J102124.9+130115 & 52764.246  & 1.007 &  0 & 0   & 30.41  & 26.46  & $-$1.52  &  0.04  & $-$0.44 \\
103 &   & 2XMM J102147.4+130850 & 52764.246  & 0.656 &  0 & 0   & 30.05  & 26.35  & $-$1.42  &  0.07  & $-$0.38 \\
104 &   & 2XMM J102350.9+041542 & 51883.770  & 1.809 &  0 & 0   & 30.93  & 26.46  & $-$1.72  & $-$0.07  & $-$0.28 \\
105 &   & 2XMM J103031.6+052455 & 52781.621  & 1.183 &  0 & 0   & 31.36  & 26.48  & $-$1.87  & $-$0.14  & $-$0.38 \\
106 &   & 2XMM J103216.0+505119 & 53473.809  & 0.173 &  0 & 0   & 29.15  & 25.63  & $-$1.35  & $-$0.02  & $-$0.11 \\
107 &   & 2XMM J103338.7+004226 & 53715.141  & 0.361 &  0 & 0   & 29.10  & 24.88  & $-$1.62  & $-$0.29  & $-$0.62 \\
108 &   & 2XMM J103413.9+585252 & 52930.035  & 0.745 &  0 & 0   & 30.78  & 26.33  & $-$1.71  & $-$0.09  & $-$0.46 \\
109 &   & 2XMM J103922.6+643417 & 53677.684  & 2.128 & $-$1 & 0   & 31.19  & 26.74  & $-$1.71  & $-$0.01  & $-$0.19 \\
110 &   & 2XMM J103935.7+533039 & 52040.297  & 0.229 &  0 & 0   & 28.79  & 25.43  & $-$1.29  & $-$0.02  & $-$0.36 \\
111 &   & 2XMM J103951.5+643005 & 53677.684  & 0.402 &  1 & 0   & 28.82  & 24.24  & $-$1.76  & $-$0.48  & $-$0.41 \\
112 &   & 2XMM J104155.7+061256 & 52777.816  & 1.478 &  0 & 0   & 30.62  & 26.01  & $-$1.77  & $-$0.17  & $-$0.46 \\
113 &   & 2XMM J104542.2+525112 & 53303.078  & 1.058 &  1 & 0   & 30.85  & 27.30  & $-$1.37  &  0.27  & $-$0.37 \\
114 &   & 2XMM J104609.8+530008 & 53303.078  & 1.179 &  0 & 0   & 30.61  & 26.41  & $-$1.61  & $-$0.02  & $-$0.27 \\
115 &   & 2XMM J104613.6+525554 & 53303.078  & 0.503 &  0 & 0   & 30.31  & 26.14  & $-$1.60  & $-$0.06  & $-$0.66 \\
116 & 1 & 2XMM J105039.5+572336 & 52562.273  & 1.447 &  1 & 0   & 30.52  & 26.66  & $-$1.48  &  0.10  & $-$0.30 \\
116 & 2 & 2XMM J105039.5+572336 & 52564.340  & 1.447 &  1 & 0   & 30.51  & 26.54  & $-$1.52  &  0.06  & $-$0.44 \\
117 &   & 2XMM J105143.8+335927 & 52407.168  & 0.167 &  0 & 0   & 29.65  & 26.04  & $-$1.39  &  0.03  & $-$0.39 \\
118 &   & 2XMM J105204.5+440152 & 52754.262  & 1.524 &  0 & 0   & 30.83  & 26.64  & $-$1.61  &  0.02  & $-$0.12 \\
119 & 1 & 2XMM J105221.0+440439 & 52754.262  & 0.968 &  0 & 0   & 30.80  & 26.06  & $-$1.82  & $-$0.19  & $-$0.47 \\
119 & 2 & 2XMM J105221.0+440439 & 52783.555  & 0.968 &  0 & 0   & 30.85  & 26.31  & $-$1.74  & $-$0.10  & $-$0.43 \\
120 & 1 & 2XMM J105224.9+441505 & 52754.262  & 0.443 &  0 & 0   & 29.48  & 26.29  & $-$1.23  &  0.16  & $-$0.37 \\
120 & 2 & 2XMM J105224.9+441505 & 52783.555  & 0.443 &  0 & 0   & 29.40  & 26.15  & $-$1.25  &  0.13  & $-$0.43 \\
121 & 1 & 2XMM J105239.6+572431 & 51661.148  & 1.112 &  0 & 0   & 30.91  & 26.67  & $-$1.63  &  0.02  & $-$0.41 \\
121 & 2 & 2XMM J105239.6+572431 & 52209.797  & 1.112 &  0 & 0   & 30.78  & 26.82  & $-$1.52  &  0.10  & $-$0.43 \\
121 & 3 & 2XMM J105239.6+572431 & 52217.285  & 1.112 &  0 & 0   & 30.78  & 26.81  & $-$1.52  &  0.10  & $-$0.36 \\
121 & 4 & 2XMM J105239.6+572431 & 52566.344  & 1.112 &  0 & 0   & 30.72  & 26.70  & $-$1.54  &  0.07  & $-$0.33 \\
121 & 5 & 2XMM J105239.6+572431 & 52568.359  & 1.112 &  0 & 0   & 30.73  & 26.75  & $-$1.53  &  0.09  & $-$0.41 \\
121 & 6 & 2XMM J105239.6+572431 & 52570.340  & 1.112 &  0 & 0   & 30.73  & 26.79  & $-$1.51  &  0.11  & $-$0.42 \\
121 & 7 & 2XMM J105239.6+572431 & 52605.977  & 1.112 &  0 & 0   & 30.74  & 26.70  & $-$1.55  &  0.07  & $-$0.40 \\
121 & 8 & 2XMM J105239.6+572431 & 52612.199  & 1.112 &  0 & 0   & 30.73  & 26.72  & $-$1.54  &  0.08  & $-$0.40 \\
121 & 9 & 2XMM J105239.6+572431 & 52614.160  & 1.112 &  0 & 0   & 30.73  & 26.75  & $-$1.53  &  0.09  & $-$0.38 \\
122 & 1 & 2XMM J105316.7+573550 & 51661.148  & 1.205 &  0 & 0   & 30.00  & 27.10  & $-$1.11  &  0.38  & $-$0.30 \\
122 & 2 & 2XMM J105316.7+573550 & 52570.340  & 1.205 &  0 & 0   & 30.21  & 26.99  & $-$1.24  &  0.28  & $-$0.30 \\
122 & 3 & 2XMM J105316.7+573550 & 52572.340  & 1.205 &  0 & 0   & 30.19  & 27.00  & $-$1.23  &  0.29  & $-$0.29 \\
122 & 4 & 2XMM J105316.7+573550 & 52605.977  & 1.205 &  0 & 0   & 30.26  & 27.04  & $-$1.23  &  0.30  & $-$0.29 \\
122 & 5 & 2XMM J105316.7+573550 & 52614.160  & 1.205 &  0 & 0   & 30.23  & 27.11  & $-$1.20  &  0.33  & $-$0.26 \\
123 &   & 2XMM J110334.7+355108 & 52044.387  & 1.200 &  0 & 0   & 30.05  & 25.89  & $-$1.60  & $-$0.11  & $-$0.61 \\
124 &   & 2XMM J111038.5+483116 & 52426.695  & 2.955 &  0 & 0   & 32.01  & 27.26  & $-$1.83  &  0.01  & $-$0.44 \\
125 &   & 2XMM J111706.4+441333 & 52408.680  & 0.144 &  0 & 0   & 29.71  & 26.06  & $-$1.40  &  0.03  & $-$0.10 \\
126 &   & 2XMM J111753.3+412016 & 53715.844  & 2.221 &  1 & 0   & 31.13  & 27.59  & $-$1.36  &  0.33  & $-$0.33 \\
127 &   & 2XMM J111830.2+402554 & 52411.293  & 0.155 &  0 & 0   & 29.92  & 25.99  & $-$1.51  & $-$0.04  & $-$0.53 \\
128 &   & 2XMM J112026.2+134024 & 51875.809  & 0.982 &  0 & 0   & 30.32  & 26.57  & $-$1.44  &  0.10  & $-$0.39 \\
129 &   & 2XMM J112048.9+133822 & 51875.809  & 0.513 &  0 & 0   & 29.64  & 25.04  & $-$1.76  & $-$0.34  & $-$0.02 \\
130 &   & 2XMM J112611.6+425245 & 51871.109  & 0.156 &  0 & 0   & 28.68  & 23.82  & $-$1.86  & $-$0.61  &  0.52 \\
131 &   & 2XMM J113109.4+311405 & 51870.762  & 0.290 &  1 & 0   & 30.30  & 26.82  & $-$1.33  &  0.21  & $-$0.36 \\
132 & 1 & 2XMM J113224.0+525157 & 53127.523  & 0.837 &  0 & 0   & 30.14  & 26.06  & $-$1.56  & $-$0.05  & $-$0.21 \\
132 & 2 & 2XMM J113224.0+525157 & 53313.078  & 0.837 &  0 & 0   & 30.01  & 25.88  & $-$1.59  & $-$0.10  & $-$0.48 \\
133 &   & 2XMM J114856.5+525426 & 53313.254  & 1.633 &  1 & 0   & 31.77  & 27.88  & $-$1.49  &  0.31  & $-$0.25 \\
134 &   & 2XMM J115838.5+435505 & 53687.480  & 1.208 &  1 & 0   & 30.33  & 26.31  & $-$1.54  &  0.00  & $-$0.20 \\
135 &   & 2XMM J115851.0+435048 & 53687.480  & 0.287 &  0 & 0   & 29.02  & 25.42  & $-$1.38  & $-$0.07  & $-$0.47 \\
136 &   & 2XMM J120504.4+352209 & 52782.684  & 2.279 &  0 & 0   & 31.55  & 26.98  & $-$1.75  &  0.01  & $-$0.30 \\
137 &   & 2XMM J120522.1+443141 & 52801.723  & 1.921 &  0 & 1   & 31.06  & 26.61  & $-$1.71  & $-$0.04  & $-$0.07 \\
138 &   & 2XMM J121342.9+025248 & 52273.520  & 0.641 &  0 & 0   & 29.65  & 25.50  & $-$1.59  & $-$0.17  & $-$0.61 \\
139 & 1 & 2XMM J121426.5+140259 & 52075.359  & 1.279 &  1 & 0   & 30.36  & 27.15  & $-$1.23  &  0.32  & $-$0.31 \\
139 & 2 & 2XMM J121426.5+140259 & 53177.242  & 1.279 &  1 & 0   & 30.40  & 27.23  & $-$1.22  &  0.34  & $-$0.30 \\
140 &   & 2XMM J121640.5+071224 & 53165.395  & 0.586 &  0 & 0   & 30.72  & 26.77  & $-$1.52  &  0.09  & $-$0.42 \\
141 &   & 2XMM J121713.1+070236 & 53165.395  & 1.203 &  0 & 0   & 30.73  & 26.59  & $-$1.59  &  0.03  & $-$0.58 \\
142 &   & 2XMM J121919.0+063926 & 52626.047  & 0.654 &  0 & 0   & 29.78  & 25.93  & $-$1.48  & $-$0.03  & $-$0.44 \\
143 &   & 2XMM J122018.4+064120 & 52460.359  & 0.286 &  0 & 0   & 29.63  & 26.46  & $-$1.22  &  0.20  & $-$0.38 \\
144 &   & 2XMM J122528.4+131725 & 53366.883  & 1.794 &  0 & 0   & 31.60  & 26.54  & $-$1.94  & $-$0.17  & $-$0.43 \\
145 &   & 2XMM J122556.1+130656 & 52456.777  & 1.350 &  0 & 0   & 30.68  & 26.50  & $-$1.60  &  0.01  & $-$0.26 \\
146 &   & 2XMM J122703.3+125402 & 53358.168  & 1.278 &  0 & 0   & 30.59  & 26.15  & $-$1.70  & $-$0.11  & $-$0.37 \\
147 &   & 2XMM J122742.9+013438 & 52083.336  & 1.279 &  0 & 0   & 30.54  & 26.54  & $-$1.53  &  0.05  & $-$0.24 \\
148 &   & 2XMM J122923.7+075359 & 52430.555  & 0.854 &  0 & 0   & 29.79  & 26.53  & $-$1.25  &  0.20  & $-$0.30 \\
149 & 1 & 2XMM J123049.7+640848 & 51685.082  & 1.040 & $-$1 & 0   & 30.35  & 26.18  & $-$1.60  & $-$0.05  & $-$0.50 \\
149 & 2 & 2XMM J123049.7+640848 & 52760.367  & 1.040 & $-$1 & 0   & 30.32  & 25.77  & $-$1.74  & $-$0.20  & $-$0.37 \\
150 & 1 & 2XMM J123054.1+110011 & 52833.379  & 0.236 &  0 & 0   & 29.95  & 26.36  & $-$1.38  &  0.10  & $-$0.42 \\
150 & 2 & 2XMM J123054.1+110011 & 53717.434  & 0.236 &  0 & 0   & 30.03  & 26.28  & $-$1.44  &  0.05  & $-$0.42 \\
150 & 3 & 2XMM J123054.1+110011 & 53721.234  & 0.236 &  0 & 0   & 30.02  & 26.30  & $-$1.43  &  0.06  & $-$0.43 \\
151 & 1 & 2XMM J123229.6+641115 & 51685.082  & 0.743 &  0 & 0   & 30.39  & 25.95  & $-$1.70  & $-$0.14  & $-$0.40 \\
151 & 2 & 2XMM J123229.6+641115 & 52760.367  & 0.743 &  0 & 0   & 30.45  & 26.26  & $-$1.61  & $-$0.04  & $-$0.42 \\
152 &   & 2XMM J123335.1+475801 & 53173.680  & 0.382 &  0 & 0   & 30.19  & 26.65  & $-$1.36  &  0.16  & $-$0.47 \\
153 &   & 2XMM J123356.1+074755 & 53161.492  & 0.371 &  0 & 0   & 29.31  & 26.13  & $-$1.22  &  0.14  & $-$0.39 \\
154 &   & 2XMM J123413.4+475352 & 53173.680  & 0.373 &  1 & 0   & 30.05  & 26.27  & $-$1.45  &  0.04  & $-$0.49 \\
155 &   & 2XMM J123508.2+393019 & 53149.184  & 0.968 &  0 & 0   & 30.03  & 25.87  & $-$1.60  & $-$0.11  & $-$0.42 \\
156 &   & 2XMM J123527.3+392824 & 53149.184  & 2.158 &  0 & 0   & 31.04  & 26.97  & $-$1.56  &  0.11  & $-$0.34 \\
157 & 1 & 2XMM J123622.9+621526 & 52047.398  & 2.587 &  0 & 0   & 30.79  & 26.81  & $-$1.53  &  0.10  & $-$0.17 \\
157 & 2 & 2XMM J123622.9+621526 & 52047.977  & 2.587 &  0 & 0   & 30.77  & 26.65  & $-$1.58  &  0.04  & $-$0.42 \\
157 & 3 & 2XMM J123622.9+621526 & 52056.293  & 2.587 &  0 & 0   & 30.55  & 26.70  & $-$1.48  &  0.10  & $-$0.35 \\
157 & 4 & 2XMM J123622.9+621526 & 52061.379  & 2.587 &  0 & 0   & 30.55  & 26.68  & $-$1.49  &  0.09  & $-$0.37 \\
157 & 5 & 2XMM J123622.9+621526 & 52967.422  & 2.587 &  0 & 0   & 30.57  & 26.02  & $-$1.75  & $-$0.16  & $-$0.84 \\
157 & 6 & 2XMM J123622.9+621526 & 52987.977  & 2.587 &  0 & 0   & 30.46  & 26.51  & $-$1.51  &  0.06  & $-$0.08 \\
158 & 1 & 2XMM J123759.5+621102 & 52047.398  & 0.910 &  0 & 0   & 30.51  & 26.65  & $-$1.48  &  0.10  & $-$0.35 \\
158 & 2 & 2XMM J123759.5+621102 & 52047.977  & 0.910 &  0 & 0   & 30.49  & 26.72  & $-$1.45  &  0.12  & $-$0.30 \\
158 & 3 & 2XMM J123759.5+621102 & 52056.293  & 0.910 &  0 & 0   & 30.53  & 26.58  & $-$1.52  &  0.06  & $-$0.39 \\
158 & 4 & 2XMM J123759.5+621102 & 52061.379  & 0.910 &  0 & 0   & 30.52  & 26.59  & $-$1.51  &  0.07  & $-$0.33 \\
158 & 5 & 2XMM J123759.5+621102 & 52967.422  & 0.910 &  0 & 0   & 30.31  & 26.43  & $-$1.49  &  0.05  & $-$0.23 \\
159 & 1 & 2XMM J123800.9+621336 & 52047.398  & 0.440 &  0 & 0   & 29.59  & 25.58  & $-$1.54  & $-$0.13  & $-$0.53 \\
159 & 2 & 2XMM J123800.9+621336 & 52047.977  & 0.440 &  0 & 0   & 29.57  & 25.48  & $-$1.57  & $-$0.16  & $-$0.55 \\
159 & 3 & 2XMM J123800.9+621336 & 52056.293  & 0.440 &  0 & 0   & 29.57  & 25.33  & $-$1.63  & $-$0.22  & $-$0.46 \\
159 & 4 & 2XMM J123800.9+621336 & 52061.379  & 0.440 &  0 & 0   & 29.55  & 25.50  & $-$1.56  & $-$0.15  & $-$0.47 \\
159 & 5 & 2XMM J123800.9+621336 & 52967.422  & 0.440 &  0 & 0   & 29.48  & 25.42  & $-$1.56  & $-$0.17  & $-$0.24 \\
159 & 6 & 2XMM J123800.9+621336 & 52979.543  & 0.440 &  0 & 0   & 29.46  & 25.33  & $-$1.58  & $-$0.19  & $-$0.32 \\
160 &   & 2XMM J124406.9+113524 & 51911.711  & 1.344 &  0 & 0   & 30.26  & 26.80  & $-$1.33  &  0.20  & $-$0.37 \\
161 &   & 2XMM J124540.9$-$002744 & 52452.605  & 1.693 &  0 & 0   & 31.07  & 27.29  & $-$1.45  &  0.23  & $-$0.33 \\
162 &   & 2XMM J124728.5+671725 & 53703.648  & 1.220 & $-$1 & 0   & 30.57  & 26.43  & $-$1.59  & $-$0.00  & $-$0.38 \\
163 &   & 2XMM J125535.1+565238 & 52067.582  & 1.803 &  0 & 0   & 30.84  & 26.33  & $-$1.73  & $-$0.09  & $-$0.52 \\
164 &   & 2XMM J125536.2+564959 & 52067.582  & 1.374 &  0 & 0   & 30.24  & 25.94  & $-$1.65  & $-$0.12  &  0.08 \\
165 &   & 2XMM J125642.1+564719 & 52067.582  & 1.956 &  0 & 0   & 30.62  & 26.50  & $-$1.58  &  0.02  & $-$0.45 \\
166 &   & 2XMM J125840.2+283426 & 51719.062  & 1.321 &  0 & 0   & 30.66  & 26.03  & $-$1.78  & $-$0.18  & $-$0.27 \\
167 &   & 2XMM J125849.8$-$014303 & 52272.977  & 0.967 &  0 & 0   & 31.13  & 26.98  & $-$1.59  &  0.10  & $-$0.44 \\
168 & 1 & 2XMM J125903.9+344702 & 52804.578  & 0.608 &  0 & 0   & 29.79  & 26.20  & $-$1.38  &  0.07  & $-$0.34 \\
168 & 2 & 2XMM J125903.9+344702 & 52976.324  & 0.608 &  0 & 0   & 29.41  & 25.94  & $-$1.33  &  0.05  & $-$0.35 \\
169 & 1 & 2XMM J130028.5+283010 & 52432.785  & 0.649 &  1 & 0   & 30.57  & 27.02  & $-$1.36  &  0.23  & $-$0.34 \\
169 & 2 & 2XMM J130028.5+283010 & 53162.824  & 0.649 &  1 & 0   & 30.73  & 27.05  & $-$1.41  &  0.21  & $-$0.35 \\
169 & 3 & 2XMM J130028.5+283010 & 53174.715  & 0.649 &  1 & 0   & 30.73  & 27.08  & $-$1.40  &  0.22  & $-$0.34 \\
169 & 4 & 2XMM J130028.5+283010 & 53198.680  & 0.649 &  1 & 0   & 30.73  & 27.11  & $-$1.39  &  0.23  & $-$0.37 \\
170 & 1 & 2XMM J130048.1+282321 & 52432.785  & 1.924 &  0 & 0   & 31.58  & 26.83  & $-$1.82  & $-$0.05  & $-$0.43 \\
170 & 2 & 2XMM J130048.1+282321 & 53162.824  & 1.924 &  0 & 0   & 31.44  & 26.66  & $-$1.84  & $-$0.10  & $-$0.35 \\
170 & 3 & 2XMM J130048.1+282321 & 53174.715  & 1.924 &  0 & 0   & 31.43  & 26.62  & $-$1.85  & $-$0.11  & $-$0.34 \\
170 & 4 & 2XMM J130048.1+282321 & 53198.680  & 1.924 &  0 & 0   & 31.43  & 26.64  & $-$1.84  & $-$0.10  & $-$0.41 \\
171 &   & 2XMM J130257.8+673006 & 52381.281  & 1.837 &  1 & 0   & 31.03  & 26.84  & $-$1.61  &  0.06  & $-$0.27 \\
172 &   & 2XMM J130454.3+673007 & 52381.281  & 0.539 & $-$1 & 0   & 29.59  & 25.80  & $-$1.46  & $-$0.05  & $-$0.37 \\
173 &   & 2XMM J130942.2$-$014139 & 53539.754  & 0.824 &  0 & 0   & 30.26  & 25.71  & $-$1.75  & $-$0.22  & $-$0.31 \\
174 &   & 2XMM J130952.0$-$013217 & 53539.754  & 1.844 &  1 & 0   & 30.57  & 26.78  & $-$1.45  &  0.14  & $-$0.26 \\
175 &   & 2XMM J131817.6+324053 & 52465.020  & 1.647 &  0 & 0   & 31.33  & 27.05  & $-$1.64  &  0.08  & $-$0.39 \\
176 &   & 2XMM J132419.8+053704 & 53197.707  & 0.203 &  1 & 0   & 28.80  & 23.65  & $-$1.98  & $-$0.71  & $-$0.54 \\
177 &   & 2XMM J132607.0+655543 & 53125.266  & 1.513 & $-$1 & 0   & 30.99  & 26.52  & $-$1.72  & $-$0.06  & $-$0.37 \\
178 &   & 2XMM J132623.0+011501 & 53370.500  & 1.232 &  0 & 0   & 30.43  & 26.12  & $-$1.65  & $-$0.09  & $-$0.32 \\
179 &   & 2XMM J132711.1+011010 & 53370.500  & 0.971 &  0 & 0   & 30.54  & 26.34  & $-$1.61  & $-$0.03  & $-$0.36 \\
180 &   & 2XMM J133526.7+405958 & 52621.199  & 1.765 &  0 & 0   & 31.16  & 26.65  & $-$1.73  & $-$0.04  & $-$0.06 \\
181 &   & 2XMM J134044.5$-$004516 & 53549.691  & 0.386 &  0 & 0   & 30.03  & 25.99  & $-$1.55  & $-$0.06  & $-$0.40 \\
182 &   & 2XMM J134113.9$-$005314 & 53549.691  & 0.237 &  1 & 0   & 29.39  & 26.41  & $-$1.14  &  0.24  & $-$0.36 \\
183 &   & 2XMM J134252.9+403202 & 52433.379  & 0.906 &  1 & 0   & 29.78  & 26.74  & $-$1.17  &  0.28  & $-$0.26 \\
184 & 1 & 2XMM J134256.5+000057 & 52482.906  & 0.804 &  0 & 0   & 30.01  & 26.64  & $-$1.30  &  0.19  & $-$0.22 \\
184 & 2 & 2XMM J134256.5+000057 & 53033.730  & 0.804 &  0 & 0   & 29.90  & 26.69  & $-$1.23  &  0.24  & $-$0.34 \\
185 &   & 2XMM J134740.9+581242 & 52066.254  & 2.050 &  1 & 0   & 31.66  & 27.64  & $-$1.54  &  0.24  & $-$0.36 \\
186 &   & 2XMM J134749.8+582109 & 52066.254  & 0.646 &  0 & 0   & 30.83  & 27.08  & $-$1.44  &  0.19  & $-$0.37 \\
187 &   & 2XMM J134834.2+262205 & 52652.574  & 0.918 &  0 & 0   & 30.42  & 26.54  & $-$1.49  &  0.07  & $-$0.59 \\
188 &   & 2XMM J134848.2+262219 & 52652.574  & 0.595 &  0 & 0   & 30.06  & 26.07  & $-$1.53  & $-$0.03  & $-$0.46 \\
189 &   & 2XMM J135639.1+051950 & 53213.418  & 1.394 &  0 & 0   & 30.49  & 26.76  & $-$1.43  &  0.14  & $-$0.46 \\
190 &   & 2XMM J140040.4+621243 & 53109.504  & 0.661 &  0 & 0   & 29.48  & 25.79  & $-$1.42  & $-$0.03  & $-$0.58 \\
191 & 1 & 2XMM J141642.3+521813 & 51746.934  & 1.285 &  0 & 0   & 31.01  & 26.38  & $-$1.78  & $-$0.11  & $-$0.28 \\
191 & 2 & 2XMM J141642.3+521813 & 51748.750  & 1.285 &  0 & 0   & 31.00  & 26.39  & $-$1.77  & $-$0.11  & $-$0.12 \\
191 & 3 & 2XMM J141642.3+521813 & 51748.926  & 1.285 &  0 & 0   & 30.98  & 26.51  & $-$1.72  & $-$0.06  & $-$0.10 \\
192 &   & 2XMM J141700.7+445606 & 52616.805  & 0.114 &  0 & 0   & 29.39  & 25.61  & $-$1.45  & $-$0.07  & $-$0.47 \\
193 &   & 2XMM J142355.5+383150 & 53528.645  & 1.205 &  0 & 0   & 30.45  & 26.32  & $-$1.59  & $-$0.02  & $-$0.71 \\
194 &   & 2XMM J142406.6+383714 & 52851.906  & 1.561 &  0 & 0   & 31.02  & 26.62  & $-$1.69  & $-$0.02  & $-$0.58 \\
195 & 1 & 2XMM J142435.9+421030 & 52848.184  & 2.218 &  0 & 0   & 31.68  & 27.37  & $-$1.65  &  0.14  & $-$0.25 \\
195 & 2 & 2XMM J142435.9+421030 & 52991.941  & 2.218 &  0 & 0   & 31.63  & 27.14  & $-$1.72  &  0.06  & $-$0.44 \\
196 & 1 & 2XMM J142455.5+421408 & 52848.184  & 0.316 &  0 & 0   & 30.29  & 26.27  & $-$1.54  & $-$0.00  & $-$0.45 \\
196 & 2 & 2XMM J142455.5+421408 & 52991.941  & 0.316 &  0 & 0   & 30.22  & 26.28  & $-$1.51  &  0.02  & $-$0.41 \\
197 &   & 2XMM J142904.3+012228 & 51753.441  & 0.420 &  0 & 0   & 29.34  & 25.87  & $-$1.34  &  0.03  & $-$0.28 \\
198 &   & 2XMM J142917.6+012059 & 51753.441  & 1.133 &  0 & 0   & 30.59  & 26.40  & $-$1.61  & $-$0.02  & $-$0.35 \\
199 &   & 2XMM J142931.5+012124 & 51753.441  & 1.518 &  0 & 0   & 30.69  & 26.46  & $-$1.62  & $-$0.01  & $-$0.55 \\
200 &   & 2XMM J142943.0+474726 & 52425.266  & 0.221 &  0 & 0   & 29.91  & 26.23  & $-$1.41  &  0.06  & $-$0.44 \\
201 & 1 & 2XMM J143025.8+415957 & 52617.281  & 0.352 &  0 & 0   & 29.26  & 25.82  & $-$1.32  &  0.03  & $-$0.43 \\
201 & 2 & 2XMM J143025.8+415957 & 52656.703  & 0.352 &  0 & 0   & 29.12  & 25.84  & $-$1.26  &  0.07  & $-$0.38 \\
202 &   & 2XMM J143424.9+033912 & 53600.020  & 1.120 &  0 & 0   & 30.61  & 26.42  & $-$1.61  & $-$0.02  & $-$0.38 \\
203 &   & 2XMM J143440.4+484139 & 52647.297  & 1.945 &  0 & 0   & 31.11  & 25.97  & $-$1.97  & $-$0.29  & $-$0.75 \\
204 &   & 2XMM J143506.5+033258 & 53600.020  & 2.404 &  0 & 0   & 30.90  & 26.85  & $-$1.55  &  0.10  & $-$0.34 \\
205 &   & 2XMM J143513.9+484149 & 52647.297  & 1.887 &  0 & 1   & 30.54  & 26.82  & $-$1.43  &  0.15  & $-$0.36 \\
206 &   & 2XMM J143621.2+484606 & 52647.297  & 2.395 &  0 & 0   & 31.15  & 26.58  & $-$1.76  & $-$0.07  & $-$0.72 \\
207 &   & 2XMM J143822.0+642000 & 51908.621  & 1.163 & $-$1 & 0   & 30.69  & 26.46  & $-$1.63  & $-$0.02  & $-$0.40 \\
208 &   & 2XMM J144414.6+063306 & 53412.715  & 0.208 &  0 & 0   & 29.64  & 26.49  & $-$1.21  &  0.21  & $-$0.38 \\
209 &   & 2XMM J144645.9+403506 & 52497.062  & 0.267 &  0 & 0   & 30.39  & 26.13  & $-$1.63  & $-$0.07  & $-$0.54 \\
210 &   & 2XMM J150121.9+014401 & 53574.242  & 0.608 &  1 & 0   & 29.18  & 26.36  & $-$1.08  &  0.26  &  0.06 \\
211 &   & 2XMM J150148.8+014403 & 53574.242  & 0.484 &  0 & 0   & 29.45  & 25.79  & $-$1.40  & $-$0.01  & $-$0.37 \\
212 &   & 2XMM J150948.6+333626 & 52852.219  & 0.512 &  0 & 0   & 29.40  & 25.45  & $-$1.52  & $-$0.14  & $-$0.46 \\
213 &   & 2XMM J151443.0+365050 & 52511.234  & 0.371 &  1 & 0   & 30.41  & 26.95  & $-$1.33  &  0.23  & $-$0.38 \\
214 &   & 2XMM J151551.6+000304 & 53209.027  & 1.775 &  0 & 0   & 30.96  & 26.29  & $-$1.79  & $-$0.13  &  0.02 \\
215 &   & 2XMM J151630.2$-$010108 & 53220.680  & 1.212 &  0 & 0   & 30.37  & 26.20  & $-$1.60  & $-$0.05  & $-$0.48 \\
216 &   & 2XMM J151630.2$-$005625 & 53220.680  & 1.921 &  1 & 1   & 30.98  & 26.77  & $-$1.62  &  0.04  & $-$0.34 \\
217 &   & 2XMM J151652.7$-$005834 & 53220.680  & 1.722 &  0 & 0   & 31.10  & 26.89  & $-$1.62  &  0.06  & $-$0.29 \\
218 &   & 2XMM J151842.8+424933 & 53245.457  & 1.465 &  0 & 0   & 30.73  & 26.52  & $-$1.62  & $-$0.00  & $-$0.14 \\
219 & 1 & 2XMM J152553.8+513649 & 52252.004  & 2.882 &  0 & 1   & 31.69  & 27.87  & $-$1.47  &  0.32  & $-$0.37 \\
219 & 2 & 2XMM J152553.8+513649 & 52256.246  & 2.882 &  0 & 1   & 31.64  & 27.84  & $-$1.46  &  0.32  & $-$0.39 \\
220 &   & 2XMM J153322.8+324351 & 52485.520  & 1.899 &  0 & 0   & 30.57  & 26.86  & $-$1.42  &  0.17  & $-$0.31 \\
221 &   & 2XMM J153438.1+553945 & 52411.594  & 1.130 &  0 & 0   & 30.26  & 26.43  & $-$1.47  &  0.06  & $-$0.50 \\
222 &   & 2XMM J154530.3+484608 & 52678.859  & 0.400 &  0 & 0   & 30.53  & 26.07  & $-$1.71  & $-$0.13  & $-$0.57 \\
223 &   & 2XMM J160106.2+084605 & 52860.309  & 1.207 &  0 & 0   & 30.47  & 27.03  & $-$1.32  &  0.25  & $-$0.35 \\
224 &   & 2XMM J160658.2+271706 & 53580.660  & 0.934 &  1 & 0   & 30.41  & 27.22  & $-$1.22  &  0.34  & $-$0.20 \\
225 & 1 & 2XMM J162855.6+394034 & 52461.664  & 1.520 &  0 & 0   & 30.74  & 26.29  & $-$1.71  & $-$0.09  & $-$0.66 \\
225 & 2 & 2XMM J162855.6+394034 & 52501.555  & 1.520 &  0 & 0   & 30.81  & 26.79  & $-$1.54  &  0.09  & $-$0.24 \\
226 &   & 2XMM J164221.1+390333 & 53237.520  & 1.713 &  0 & 0   & 30.95  & 26.78  & $-$1.60  &  0.06  & $-$0.34 \\
227 &   & 2XMM J165713.2+352441 & 51972.938  & 2.329 &  0 & 0   & 31.12  & 26.93  & $-$1.61  &  0.08  & $-$0.16 \\
228 &   & 2XMM J170100.6+641208 & 52425.770  & 2.736 & $-$1 & 0   & 32.32  & 27.52  & $-$1.84  &  0.06  & $-$0.43 \\
229 & 1 & 2XMM J170639.3+240606 & 52707.043  & 0.836 &  0 & 0   & 29.67  & 25.90  & $-$1.45  & $-$0.02  & $-$0.32 \\
229 & 2 & 2XMM J170639.3+240606 & 52864.641  & 0.836 &  0 & 0   & 29.56  & 26.01  & $-$1.36  &  0.05  & $-$0.13 \\
230 &   & 2XMM J171359.4+640939 & 52615.277  & 1.359 &  0 & 0   & 30.80  & 26.50  & $-$1.65  & $-$0.02  & $-$0.58 \\
231 &   & 2XMM J212909.6+001214 & 52576.086  & 1.339 &  0 & 0   & 30.39  & 25.93  & $-$1.71  & $-$0.15  & $-$0.58 \\
232 & 1 & 2XMM J221738.4+001207 & 52596.160  & 1.121 &  0 & 0   & 29.81  & 25.84  & $-$1.52  & $-$0.07  & $-$0.40 \\
232 & 2 & 2XMM J221738.4+001207 & 52623.918  & 1.121 &  0 & 0   & 29.86  & 25.85  & $-$1.54  & $-$0.08  & $-$0.39 \\
233 & 1 & 2XMM J221751.3+001146 & 52596.160  & 1.491 &  0 & 0   & 30.38  & 26.36  & $-$1.55  &  0.00  & $-$0.37 \\
233 & 2 & 2XMM J221751.3+001146 & 52623.918  & 1.491 &  0 & 0   & 30.39  & 26.36  & $-$1.55  &  0.01  & $-$0.39 \\
234 & 1 & 2XMM J221755.2+001513 & 52596.160  & 2.092 &  0 & 0   & 30.47  & 26.37  & $-$1.57  & $-$0.00  & $-$0.39 \\
234 & 2 & 2XMM J221755.2+001513 & 52623.918  & 2.092 &  0 & 0   & 30.32  & 26.52  & $-$1.46  &  0.08  & $-$0.29 \\
235 &   & 2XMM J223607.6+134355 & 52787.641  & 0.326 & $-$1 & 0   & 30.41  & 26.31  & $-$1.57  & $-$0.01  & $-$0.50 \\
236 &   & 2XMM J233706.4+002132 & 52249.578  & 0.713 &  0 & 0   & 29.37  & 26.10  & $-$1.25  &  0.12  & $-$0.34 \\
237 &   & 2XMM J233707.2+002007 & 52249.578  & 1.901 &  0 & 0   & 31.11  & 26.70  & $-$1.69  & $-$0.01  & $-$0.58 \\
238 &   & 2XMM J233722.0+002238 & 52249.578  & 1.376 &  0 & 0   & 30.46  & 26.00  & $-$1.71  & $-$0.14  & $-$0.51 \\
239 &   & 2XMM J234715.3+005808 & 52636.617  & 1.487 &  0 & 0   & 30.22  & 26.66  & $-$1.37  &  0.16  & $-$0.48 \\
240 &   & 2XMM J234715.9+005602 & 52636.617  & 0.456 &  0 & 0   & 29.04  & 25.63  & $-$1.31  &  0.01  & $-$0.50 \\
241 &   & 2XMM J234724.7+005248 & 52636.617  & 1.323 &  1 & 0   & 30.53  & 26.98  & $-$1.36  &  0.22  & $-$0.28 \\
\end{longtable}
\noindent
$^{\rm a}$ 1=radio-loud; 0=radio-quiet; -1=radio-unclassified \\
$^{\rm b}$ 1=BAL; 0=non-BAL
}
\end{document}